\documentclass[useAMS,usegraphicx,usenatbib]{mn2e}
\usepackage{psfig}
\usepackage{graphicx}
\usepackage{amsmath}
\def\ltsima{$\; \buildrel < \over \sim \;$}
\def\lsim{\lower.5ex\hbox{\ltsima}}
\def\gtsima{$\; \buildrel > \over \sim \;$}
\def\gsim{\lower.5ex\hbox{\gtsima}}

\title[Fallback discs in neutron stars]{The influence of fallback discs on the spectral and timing properties of neutron stars}
\author[Yan, Perna \& Soria]{Ting Yan$^1$\thanks{E-mail: ting.yan@colorado.edu (TY); 
rosalba@colorado.edu (RP)}, 
Rosalba Perna$^{1,2}$, Roberto Soria$^3$\\
$^1$Department of Astrophysical
and Planetary Sciences, University of Colorado, Boulder, CO 80309, USA \\
$^2$JILA, Boulder, CO, 80309, USA \\
$^3$International Centre for Radio Astronomy Research, Curtin University, 
GPO Box U1987, Perth, WA 6845, Australia
}

\begin{document}

\date{Accepted 2012 April 4. Received 2012 March 19; in original form 2012 January 23}

\pagerange{\pageref{firstpage}--\pageref{lastpage}} \pubyear{2012}

\maketitle

\label{firstpage}

\begin{abstract}
Fallback discs around neutron stars (NSs) are believed to be an expected
outcome of supernova explosions. Here we investigate the consequences
of such a common outcome for the timing and spectral properties of the
associated NS population, using Monte Carlo population
synthesis models. We find that the long-term torque exterted by the
fallback disc can substantially influence the late-time period
distribution, but with quantitative differences which depend on
whether the initial spin distribution is dominated by slow or fast
pulsars.  For the latter, a single-peaked initial spin distribution
becomes bimodal at later times. Timing ages tend to understimate the
real age of older pulsars, and overestimate the age of younger ones.
{Braking indices cluster in the range $1.5\lsim n\lsim 3$ for
  slow-born pulsars, and $-0.5\lsim n\lsim 5$ for fast-born pulsars,
  with the younger objects found predominantly below $n\lsim 3$.
  Large values of $n$, while not common, are possible, and associated
  with torque transitions in the NS+disc system. The 0.1-10~keV thermal
luminosity of the NS+disc system is found to be generally dominated by
the disc emission at early times, $t\la 10^3$~yr, but this declines
  faster than the thermal surface emission of the neutron
  star}. Depending on the initial parameters, there can be occasional
periods in which some NSs switch from the propeller to the
accretion phase, increasing their luminosity up to the Eddington limit
for $\sim 10^3$--$10^4$ years.

\end{abstract}


\section{Introduction}

Neutron stars (NSs) are born in supernova (SN) explosions. While 
the stellar core collapses, the envelope is ejected outwards. 
The ejected material carries angular momentum, due to the rotation 
of the progenitor star. If a fraction of the ejecta 
does not have enough energy to escape, it will form a 
fallback accretion disc around the collapsed core 
\citep{mic88,che89}. Although a systematic and
quantitative study of the fate of the ejected material following
a SN explosion is still lacking, fallback discs around NSs
have been invoked to explain a wide range of NS properties, such as
enhanced X-ray luminosities in the Anomalous X-ray Pulsar 
\citep{cha00,cha00b,alp01a,alp01b}, pulsar braking indices that suggest 
non-dipolar torques \citep{men01a}, pulsar jets (Blackman \& Perna 2004;  
Blaschke, Grigorian \& Voskresensky 2004), 
discrepancies between the real (kinematic) pulsar ages
and their characteristic ages $\tau=P/2\dot{P}$ \citep{mar01a,mar01b}.

From an observational point of view, the search for fallback discs has
been quite active in the last decade or so. Following up on the
suggestion that fallback discs around NSs are brighter in the infrared (IR) and
at longer wavelengths \citep{per00a,per00b}, a
number of isolated NSs have been imaged in those bands 
\citep{coe98,hul00a,hul00b,kap01,isr03,mig07,wan07a,wan07b,pos10}. 
In particular, \citet{wan06} 
reported the first direct detection of a fallback disc around 
an isolated NS, based on an IR excess which could be well 
modelled with emission from a passive, dusty disc.  
Evidence for IR excess has also been reported in other
NS systems \citep{isr03,kap09}.

If fallback discs around NSs are indeed a ubiquitous phenomenon, they
can substantially interfere with the spin evolution of the NS. 
This will  affect estimates of the NS spin birth 
distribution based on the measured values of the period $P$ and 
its derivative $\dot{P}$, and the simple assumption 
that the pulsar slow down is due to dipole radiation only. 
The distribution of NS spin births, which
contains information on the physics of the SN explosion, has indeed
been the subject of of extensive investigation for several decades
\citep{gun70,phi81,sri84,lyn85,nar90,arz02,fau06,per08}.

In this paper we perform a statistical study of the spin
evolution of NSs under the combined evolution resulting from
dipole radiation losses and the interaction with a fallback disc.  We
aim at a) assessing to what extent the inferred spin birth
properties of pulsars can be influenced by the presence of fallback
discs, b) identifying observational properties that can help
confirm/rule out a ubiquitous presence of fallback discs and c)
exploring whether fallback discs can help explain some currently
puzzling pulsar observations.

Our paper is organized as follows: the coupled evolution of the disc
and pulsar system is described in Section 2.1. Evolutionary tracks for
several specific combinations of initial conditions are presented and
discussed in Section 2.2 (timing properties) and Section 2.3 (luminosity).  We
present the statistical timing and luminosity properties of the pulsar
population, derived via Monte Carlo simulations, in Section 3.1, while
  in Section 3.2 we further discuss the luminosity properties of the NS+disc
  system, and its differences from other astrophysical sources.
We summarize and conclude in Section 4, highlighting puzzling pulsar observations
which can be accounted for by the presence of fallback discs.

\section{A model of fallback accretion}

\subsection{Time evolution of the accretion flow}

We model the combined dipole/disc torque acting on the NS (rotating at angular velocity $\Omega$) 
following the approach of \citet{men99} \citep[see also][]{men01a}: 
\begin{equation}
I \dot{\Omega} = -\beta \Omega^3 + 
2 \dot{M} R_{\rm in}^2 \Omega_{\text{K}} (R_{\rm in}) \left[ 1- \frac{\Omega}
{\Omega_{\text{K}}(R_{\rm in})} \right],
\label{eq:torque}
\end{equation}
{where $R_{\rm in}$ is the inner radius of the disc}, $I=
\frac{2}{5} M_{\text{NS}} R_{\text{NS}}^2$ is the moment of inertia of
the NS, $\beta = {B^2 R_{\text{NS}}^6}/{6c^3}$ (assuming for
simplicity an orthogonal rotator), $\dot{M}$ is the accretion rate and
$\Omega_{\text{K}}$ is the Keplerian angular velocity, both evaluated
at $R_{\rm in}$.  The first term in the r.h.s. of Eq.(\ref{eq:torque})
represents the torque due to dipole radiation losses, whereas the second
term describes the torque exerted by the disc.
At high accretion rates, the ram pressure of the
accreting material causes the disc to penetrate inside the
magnetosphere, up to an inner radius at which the magnetic pressure
balances the ram pressure. This special location is called the
magnetospheric radius, $R_{\rm m} = 2.55 \times 10^8 \dot{M}^{-2/7}_{16}
M^{-1/7}_{\text{NS},1} B^{4/7}_{12} \, \text{cm}$.

{If, at any point during the coupled NS+disc evolution, 
the magnetospheric radius, $R_{\rm m}$, formally exceeds the light cylinder radius,
$R_{\text{lc}}\equiv c/\Omega$, the field lines on the disc are no longer able to exert
a significant pressure on the disc. On the other hand, the ram pressure
of the material from the accreting disc continues to act. The disc is not
able to penetrate inside $R_{\text{lc}}$, because the magnetic pressure from the
NS would then push it outwards. Therefore, when $R_{\rm m}$ becomes formally
larger than $R_{\text{lc}}$, we assume that the disc torque is applied at $R_{\text{lc}}$.  
In summary, the inner disc radius is located at $R_{\rm in}=\min\left[R_{\rm m}, R_{\text{lc}}\right]$}.

Inspection of Eq.(\ref{eq:torque}) shows that, at any given time, if
$\Omega<\Omega_{\rm K}(R_{\rm in})$, the disc exterts a positive torque,
which, if dominant over the dipole component, would make the pulsar
spin up and accrete. On the other hand, if $\Omega>\Omega_{\rm
  K}(R_{\rm in})$, material from the disc is stopped by the centrifugal
barrier \citep{ill75}, and the disc contributes to furher spin down
the pulsar.

{Formation of the fallback disc ensue as
material from the stellar envelope is thrown out at large distances
during the supernova explosion; 
the fraction of this material which remains bound, falls back, and
circularizes onto Keplerian orbits corresponding
to their original specific angular momentum (Chevalier 1989; MacFadyen et al.
2001; see also Perna \& MacFadyen 2010 for an exploration of different
angular momenta distributions in collapsing stars). }

Once the initial disc has been formed, its subsequent evolution can be
modelled following the results of the numerical simulations of \citet{can90}.  
They showed that the accretion rate in the debris disc
follows the evolution
\begin{equation}
\dot{M}(t) = \left\{\begin{array}{ll}
\dot{M}_0, \; & {\rm for}\ 0 < t < t_0, \\
\dot{M}_0 \left(t/t_0\right)^{-\alpha}, \; & {\rm for}\ t > t_0,
\end{array} \right.
\label{eq:dotM}
\end{equation}
where $\dot{M}_0$ is the accretion rate during an initial plateau
phase lasting a timescale $t_0$. \citet{men01b} estimated 
the timescale of the transient accretion phase to be on the order of
\begin{equation}
t_0=\frac{R^2\Omega_{\rm K}}{{\cal{\alpha'}} c_{\rm s}^2}\approx 6.6\times 10^{-5}
(T_{\rm c,6})^{-1} R_{\rm d,8}^{1/2}(t_0)\; {\rm yr}\,, 
\label{eq:t0}
\end{equation}
where $c_{\rm s}$ is the sound speed, $\cal{\alpha'}$ the viscosity parameter, $R_{\rm d,8}(t_0)$ the initial disc radius in units of
$10^8$ cm, and $T_{\rm c,6}$ a typical disc temperature during the
early phases, in units of $10^6$ K.  
The index $\alpha$ of the powerlaw decay, was found to be $\approx 19/16$. 
A similar behaviour, i.e. a plateau followed by a powerlaw decline, was
also found in the simulations by \citet{mac01}, although
with $\alpha\approx 5/3$.  Here, we adopt $\alpha = 19/16$ for our model.

{For an estimate of the expected range of values of the initial
accretion rate, we rely on results of numerical simulations of massive
star collapse and subsequent fallback onto the compact object.
Mineshige et al. (1997) found that the fallback material accretes towards
the central object with a supercritical accretion rate, exceeding the
Eddington limit by at least a factor of $10^6$.
The simulations of \citet{mac01}, for the
case of a 25 $M_\odot$ star, yield $\dot{M}_0\sim 10^{27}-10^{29}$~g~s$^{-1}$.
More recently, fallback in supernovae was studied by Zhang, Woosley \& Heger (2008)
for a wide range of masses, and different metallicities and explosion energies.
The full hydrodynamical study in one representative case yielded initial
accretion rates of $\sim 10^{29}$~erg~s$^{-1}$. An analysis of SN~1987A (Chevalier 1989) showed that
the accretion rate at the time of the reverse shock ($7\times 10^3$~s) is $2.2\times 10^{28}$~g~s$^{-1}$.
In general, when considering a distribution of NSs, a distribution
of initial\footnote{Note that here we speak of 'initial' disc accretion rate as if the disc
had been formed instantaneously, since the timescale of disc formation,
on the order of the dynamical timescale of collapse of the star envelope (tens
of seconds), is much shorter than the timescale (thousands of years)
of interest here for the fallback disc to affect the pulsar timing
evolution.}  accretion rates is expected, as this depends on the angular
momentum of the progenitor star, its mass, and the energetics of the explosion.}
In our work, to be broadly consistent with the result of previous numerical
work, as well as of previous semianalytical investigations which used
fallback discs, we select the initial accretion rate $\dot{M}_0$ to be
distributed log-uniformly from $10^{25}\, \text{g s}^{-1}$ to
$10^{28}\, \text{g s}^{-1}$.  However, with these values, 
the mass accretion inflow at early times would be highly supercritical; 
that is, it would produce a highly super-Eddington luminosity 
if all the inflowing matter was efficiently accreted through the
disc and onto the NS. The classical Eddington limit for a NS, 
$L_{\rm Edd} = 1.3 \times 10^{38} \left(M_{\rm NS}/M_{\odot}\right) 
\approx 2 \times 10^{38}$ erg s$^{-1}$, requires only accretion rates 
$\dot{M} \sim 10^{18}$ g s$^{-1}$, for a radiatively efficient flow.

Faced with a highly supercritical accretion flow at the outer boundary, 
an accreting system has only two options \citep{pou07}: 
it can accrete most of it in a radiatively 
inefficient way ({\it e.g.}, the ``Polish donut'' or ``slim disc'' classes 
of accretion flows: \citealt{abr88,wat00}); or 
it can spend most of the accretion power to blow away the excess 
infalling mass, in a radiatively driven wind \citep{sha73}.
For an accreting black hole, observational data and hydrodynamical 
simulations suggest that the physical situation is a combination 
of both effects \citep{ohs11}.
For an accreting NS, the presence of a hard surface prevents 
advective solutions: the excess gravitational energy released 
during accretion cannot be carried through a horizon and will always be
 dissipated, eventually. Therefore, in our model we assume that 
super-critical inflows on a young NS will lead to massive outflows. 
Following \citet{sha73} and \citet{pou07}, we assume 
that only $\dot{M} \approx \dot{M}_{\rm Edd} \equiv L_{\rm Edd}/0.1c^2$ 
reaches the inner edge of the accretion disc (from where it can then 
plunge onto the NS), regardless of the mass supply at the outer disc; 
and that the accretion rate through each radius decreases linearly, 
$\dot{M} \sim R$, because of radiation-driven wind losses.
Thus, we parameterize the radial dependence of the mass flow in the disc 
at a fixed time as
\begin{equation}
\dot{M}(R) = \dot{M}(R_{\rm in}) + [\dot{M}(R_{\rm out})-\dot{M}(R_{\rm in})]
\frac{R-R_{\rm in}}{R_{out}-R_{\rm in}}\,,
\label{eq:mdotR}
\end{equation}
where $\dot{M}(R_{\rm out},t) \equiv 
\dot{M}(R_{\rm out},t_0) $, for $0 < t < t_0$, and $\dot{M}(R_{\rm
  out},t_0)\left(t/t_0\right)^{-\alpha}$, for $ t > t_0$, 
is the rate of fallback mass inflow 
at the outer boundary, and 
$\dot{M}(R_{\rm in},t)=\min\left[\dot{M}_{\rm Edd}, \dot{M}(R_{\rm out},t)\right]$. 
In the rest of this paper, for clarity, we shall use the term 
``accretion rate'' when referring to $\dot{M}(R_{\rm in})$ (what is 
actually accreted through the whole disc), and ``inflow rate'' 
for $\dot{M}(R)$ at larger radii, most of which is lost in the outflow 
at early epochs. 
(See also \citealt{kun07} for analytical modelling 
of accretion discs with a power-law scaling of the 
inflow rate $\dot{M}(R) \sim R^{b}$.)
Recalling the definition of magnetospheric radius $R_{\rm m}$, 
it is possible that at early times, when the accretion rate is 
near Eddington, $R_{\rm m} < R_{\text{NS}}$: in that case the disc extends 
all the way to the NS surface, and we assume the disc torque to be 
$\dot{M}_{\text{Edd}}R_{\text{NS}}^2\Omega_{\text{K}}(R_{\text{NS}})$.

At intermediate epochs, when $\dot{M}(R_{\rm out},t) \approx 
\dot{M}(R_{\rm in},t) < \dot{M}_{\rm Edd}$, the wind subsides, and 
the inflow rate becomes constant thoughout the disc 
and equal to the accretion rate at the inner edge of the disc.
Finally, at late times, the accretion rate declines further, until 
it reaches a value comparable to what is expected from accretion 
from the interstellar medium. From that epoch onwards, 
we take $\dot{M}$ to be constant, $\dot{M} \sim 2 \times 10^{11}$ g s$^{-1}$, 
equal to the Bondi accretion rate \citep{bon52,fra02} 
for a medium with characteristic temperature $kT \sim 10^{3}$ keV 
and number density $n \sim 1$ cm$^{-3}$. Although the precise value 
of the Bondi accretion rate in a low density medium 
depends on the details of the accretion mode \citep{bla93,pop00,per03}, 
our results, which focus on moderately young pulsars,
are not very sensitive to the choice of the late-time asymptotic value.

\subsection{Time evolution of the pulsar timing properties}

As discussed above, at any time during the pulsar evolution, the disc can
either contribute to spinning down the pulsar even more (with respect to the
dipole term alone), or it can offset the dipole term and spin the pulsar
up. Fig.~1 shows, for several values of the accretion rate, the $P-B$ parameter
space in which the pulsar spins up (down). 

\begin{figure}
\label{fig:PB}
\centering
\includegraphics[width=8.7cm]{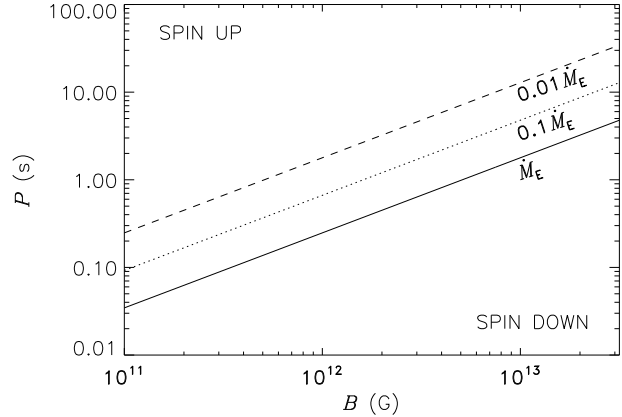}
\caption{Range of the $P-B$ parameter space in which the pulsar is either 
spinning up or down, owing to the combined effect of dipole radiation losses 
and torque by a disc. For each value of the accretion rate 
at the magnetospheric radius, the region above its corresponding curve 
represents the parameter space where the pulsar is spun up, while 
the region below represents spin down conditions.}
\end{figure}

At late times,  the torque may disappear if the disc becomes neutral, viscous dissipation 
drops and accretion stops. \citet{men01b} estimated a neutralization timescale
\begin{equation}
  t_n=\left[\frac{R^3_{d,10}(t_0) }{\dot{M}(t_0)/10^{16} \text{\,g\,s}^{-1}}\right]^{\frac{1}{2.75}}t_0,
\label{eq:tn}
\end{equation}
where $R_{d,10}$ is the disc radius in units of $10^{10}\,\text{cm}$.
The possible onset of a transition to a phase of 'dead' disc is rather
uncertain, as it depends on the nature of viscosity; opacity (and
hence disc composition) also plays an important role, as well as the
disc size (Menou et al. 2001b). Furthermore, it has been pointed out
that irradiation by the pulsar is likely to keep the disc ionized and
continue the mass inflow (Alpar et al. 2001). Here, to avoid introducing highly
uncertain parameterizations, we assume that the disc is active throughout 
the active lifetime of the pulsar. If the disc were to become dead at some
point, from that time on the pulsar would simply continue its evolution
as an isolated pulsar, spinning down by dipole radiation losses alone
(neglecting gravitational energy losses and other secondary effects).

The birth distribution of surface magnetic fields and spin periods has been the
subject of numerous studies. Usually, a log-Gaussian distribution 
is assumed for both. For the $B$ field distribution, 
there is now a general consensus about the initial parameters; 
following \citet{arz02}, we take
\begin{equation}
\langle\log B_0(\text{G})\rangle = 12.35,\;\;\;  \sigma_{\log B_0} = 0.4\,.
\label{eq:Bdist}
\end{equation}
Whether accretion leads to field decay, and the timescale over which
the magnetic field re-emerges, are still unsettled issues, especially
with regards to quantitative details. Several papers have been devoted
to this topic (e.g. Cumming et al. 2001; Geppert et al. 1999), and the
coupled evolution between a NS with a fallback disc and a field
gradually buried by the disc accretion has been studied by Liu \& Li
(2009), using an emphirical expression for $B$-field decay presented
by Shibanazi et al. (1989), $B={B_0}/({1+\Delta M/m_B})$, where
$\Delta M$ is the accreted mass, and $m_B$ is a mass normalization
constant with typical values between $\sim 10^{-5} $ and $10^{-4}
M_{\sun}$. The resulting evolutionary tracks for the periods of the
NS studied were found to be very model-dependent, with the magnitude
of the effect dependent on both the initial $B$ field and the disc
properties. In the current work, to avoid results dependent on a
rather uncertain parametrization, we consider the case of negligible
field burying. We stress that our investigation is aimed at capturing
features of pulsar properties which could be attributed to NS-disc 
interaction, but we do not make any attempt to quantitatively model data. 

For the birth period distribution, on the other hand, there is a
widely different range of values inferred by different investigations: 
from a distribution peaking around a few ms \citep{arz02}
\begin{equation}
\langle\log P_0(\text{s})\rangle = -2.3,\;\;\; \sigma_{\log P_0} > 0.2,
\label{eq:PdistS}
\end{equation}
to one peaking at several hundred ms (Faucher-Giguere \&
Kaspi 2006)
\begin{equation}
\langle\log P_0(\text{s})\rangle = -0.5, \;\;\;\sigma_{\log P_0} = 0.2,
\label{eq:PdistL}
\end{equation}
A lower limit of several tens of ms was derived 
from the scarcity of bright X-ray pulsars associated with 
historical supernovae \citep{per08}. It should be pointed out, 
however, that all those investigations assumed that pulsar 
evolution occurs with no contribution from fallback torques.  
Therefore, we use them here as an initial reference point 
to study the effect that hypothetical fallback discs would have 
on the pulsar evolution, at a statistical level.  
We take an initial spin distribution that encompasses 
the broad range of values inferred by previous investigations.
In fact, one of the goals of our study is to uncover differences
in the effect that a torque by a disc has on NSs born with
fast and slow spins. However, we do need to note that it is
possible that discs might be much more common around slow-born NSs
than fast-born NSs, since the rotational energy of the latter
can rapidly disrupt the disc (Eksi, Hernquist \& Narayan 2005).  

\begin{figure*}
\includegraphics[width=7.8cm]{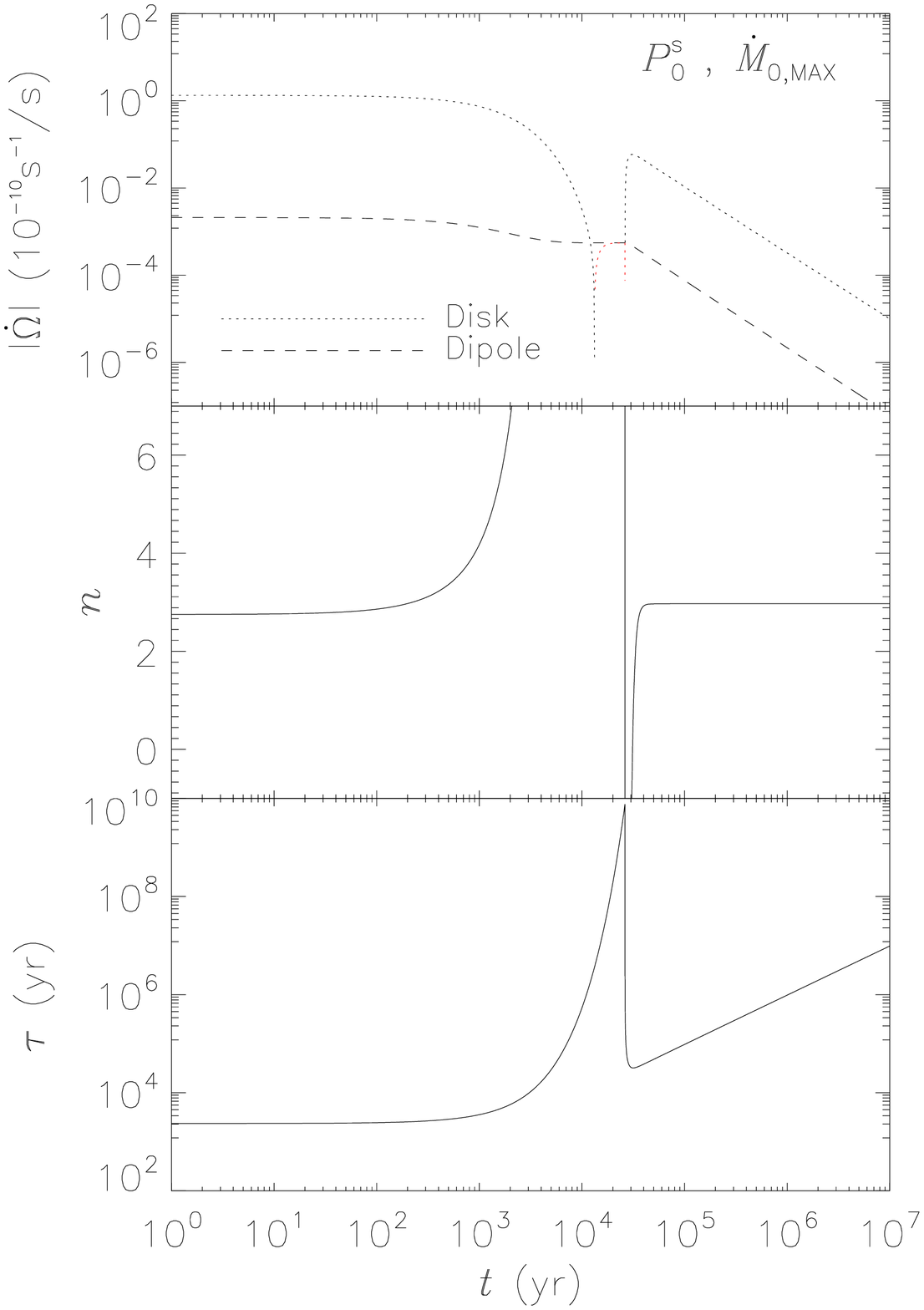}
\includegraphics[width=7.8cm]{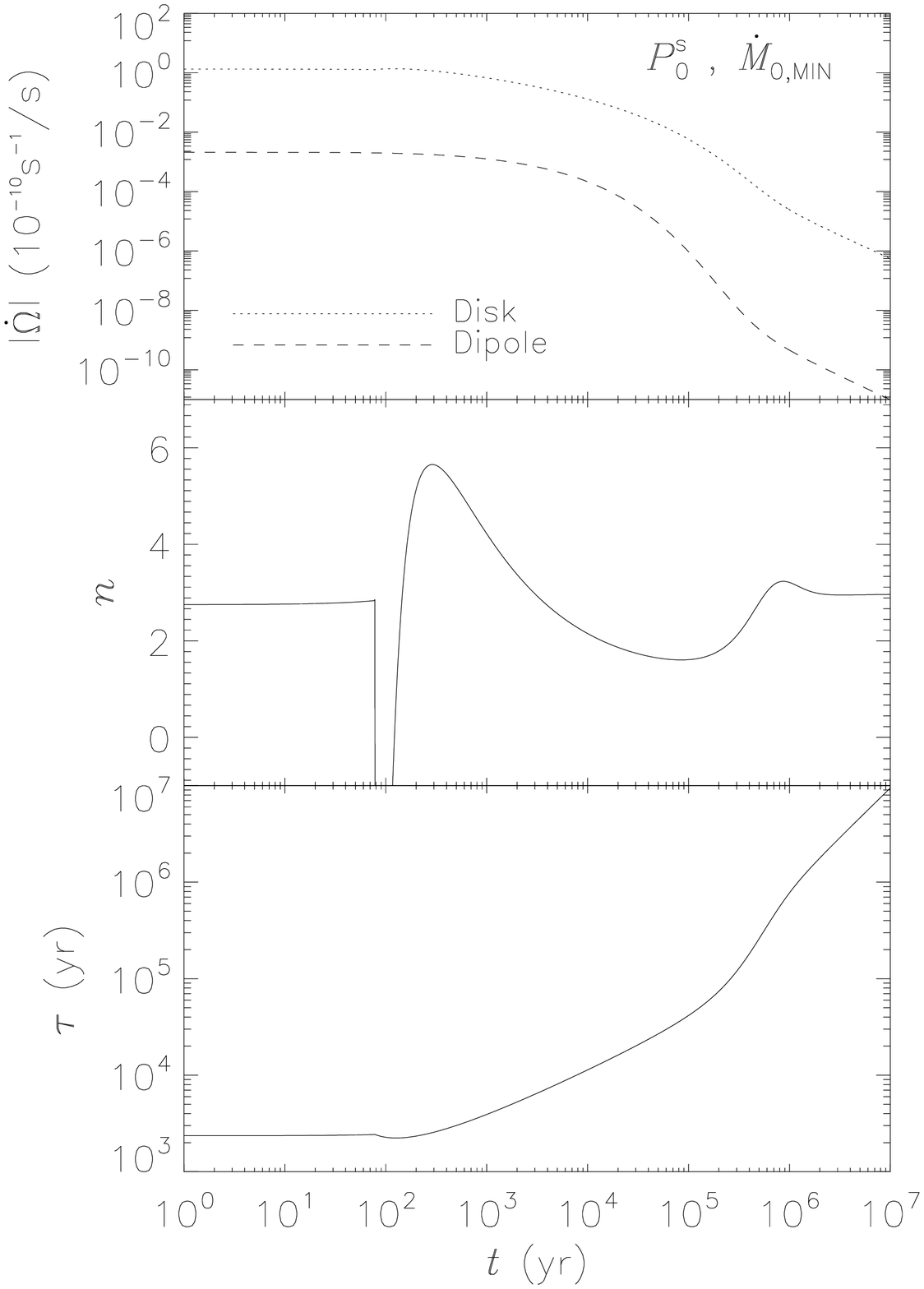}\\ 
\includegraphics[width=7.8cm]{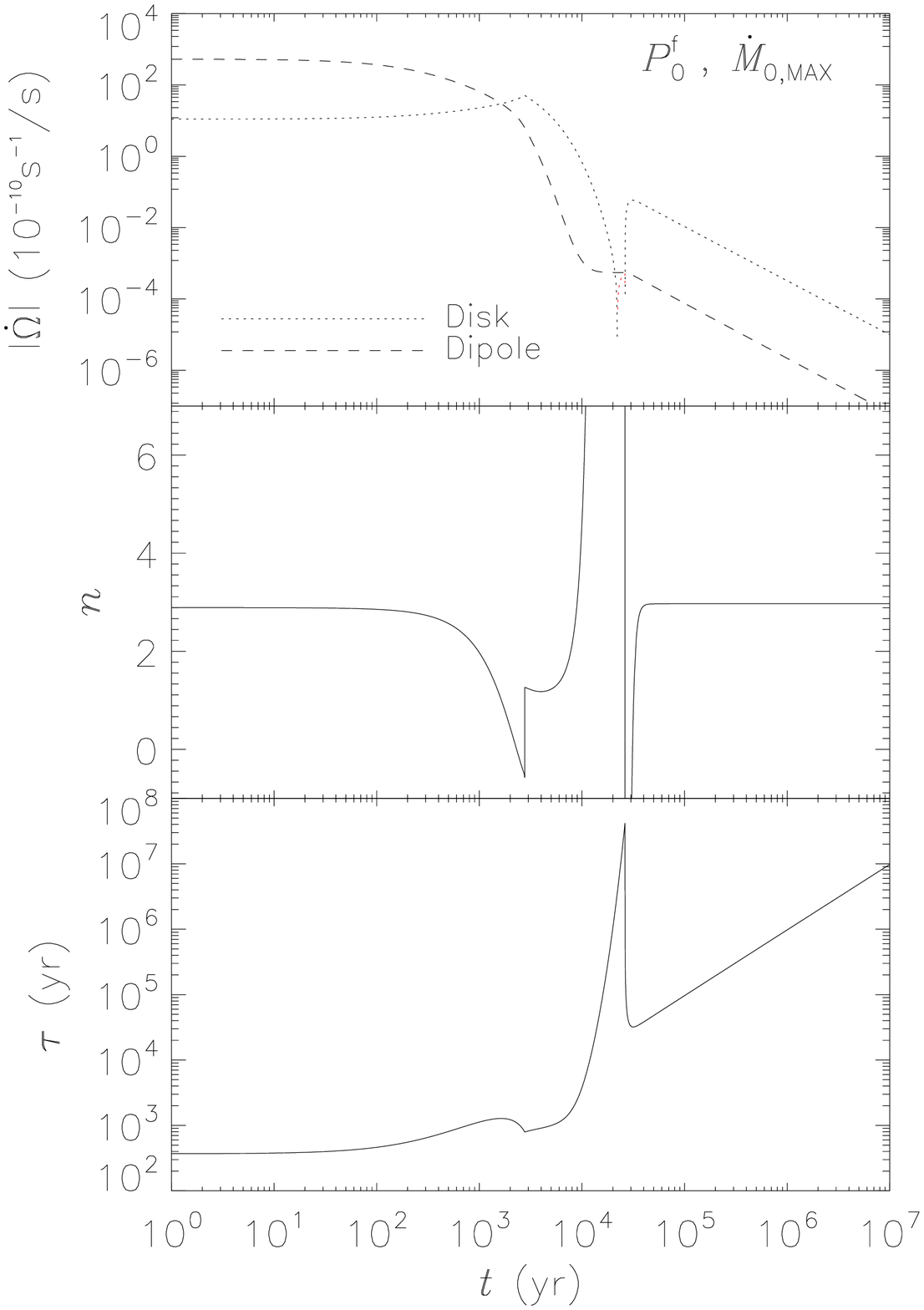}
\includegraphics[width=7.8cm]{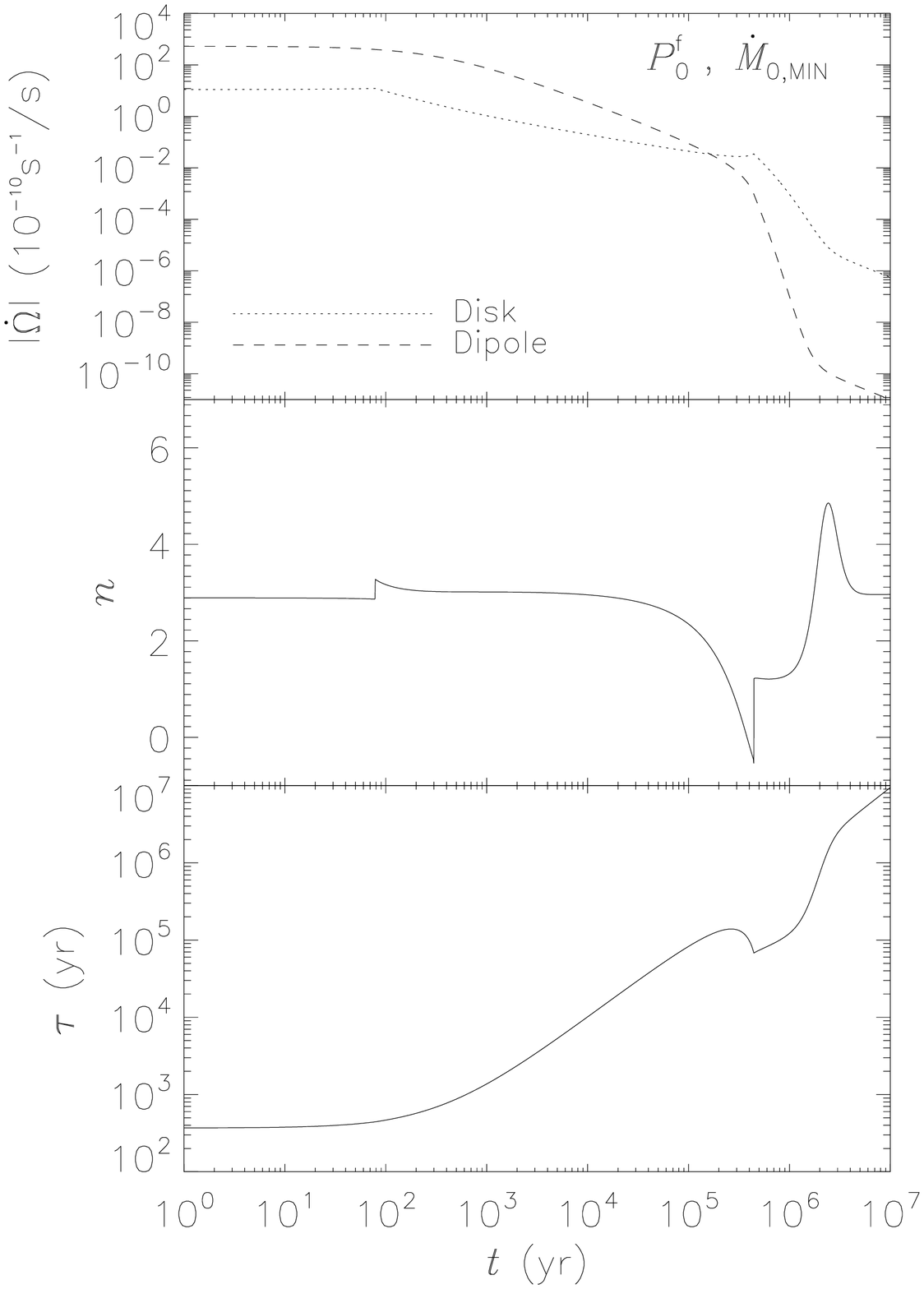}
\vspace{-0.1in}
\caption{Time evolution of the dipole and disc components of the 
spin-down rate $\left|\dot{\Omega}\right|$, of the braking index $n$, 
and of the timing age $\tau$,  for slow-born ($P_0^s=300$~ms; top panels)
and fast-born ($P_0^f=5$~ms; bottom panels) pulsars, starting 
with either a high ($\dot{M}_{0,\text{MAX}}=10^{28}$g~s$^{-1}$; left-hand panels), 
or a low ($\dot{M}_{0,\text{MIN}}=10^{25}$g~s$^{-1}$; right-hand panels) 
fallback inflow rate at the outer boundary of the disc. 
$\dot{\Omega}$ is always negative (spin-down), except for occasional brief disc spin-up 
phases, marked by red dots (left-hand panels). 
For the breaking index, we note a characteristic $n \approx 1$ transitional 
phase that happens only in fast-born pulsars (bottom panels).} 
\label{fig:ntime}
\end{figure*}

The presence of a fallback disc torque, in addition to the dipole
contribution, can be tested with measurements of the braking index,
an observable quantity defined as
\begin{equation}
n\; = \;\frac{\Omega \ddot{\Omega}}{\dot{\Omega}^2}\;,
\label{eq:braking}
\end{equation}
and of the timing age,
\begin{equation}
\tau\;=\frac{P}{2\dot{P}}\,,
\label{eq:tau}
\end{equation}
where $P=2\pi/\Omega$.
If the primary source of spin-down is due to dipole radiation losses, then $n=3$,
and the timing age of the pulsar is equal to the real age, if $P_0\ll P$.

Figure~\ref{fig:ntime} shows the time evolution of the braking index, timing
age,  and the absolute magnitude of the dipole and disc torques, 
for four representative pulsars, 
suitably chosen to represent a wide range of initial birth parameters. 
For a fixed initial $B$ field strength $B_0$
(defined in Eq~\ref{eq:Bdist}), we consider the time-evolution 
for a typical slow-born ($P_0^s=300$~ms) and fast-born
($P_0^f=5$~ms) pulsar, with an initial fallback rate either at the
maximum ($\dot{M}_{0,\text{MAX}} = 10^{28}$g~s$^{-1}$) or at the minimum value 
($\dot{M}_{0,\text{MIN}}=10^{25}$g~s$^{-1}$) of the assumed  
distribution. 
{In order to compute the evolutionary tracks shown in
  Figure~\ref{fig:ntime} by means of Eq.(\ref{eq:braking}) and
  (\ref{eq:tau}), we use Eq.(\ref{eq:torque}) to evaluate
  $\dot{\Omega}$, from which $\ddot{\Omega}$ and $\Omega$ are
  derived through numerical differentiation and numerical
  integration, respectively.  For the latter, the initial condition is
  $\Omega(t=0)=2\pi/P_0$, with $P_0=P_0^f$ or $P_0=P_0^s$ depending on
  the case.  Note that in Eq.(\ref{eq:torque}), $\dot{M}$ represents
  the accretion rate at the inner disc boundary, i.e. $\dot{M}(R_{\rm
    in})=\min\left[\dot{M}_{\rm Edd}, \dot{M}(R_{\rm out},t)\right]$,
  where, recalling from Section 2.1, $\dot{M}(R_{\rm out},t) \equiv
  \dot{M}(R_{\rm out},t_0) $, for $0 < t < t_0$ and $\dot{M}(R_{\rm
    out},t_0)\left(t/t_0\right)^{-\alpha}$ for $ t > t_0$. For each
combination of system parameters $B_0$, $P_0$, $\dot{M}(R_{\rm
    out},t_0)$, the timing evolution of the NS driven by 
the coupled disc+dipole torques is therefore uniquely determined. 

The result of our calculations is that the disc torque is always
negative, that is $\Omega > \Omega_{\rm K}$ (corresponding to 
a spin down/propeller regime), except for a few brief phases when 
the accretion rate is very high and $\Omega < \Omega_{\rm K}$ 
(corresponding to a spin up/accretor regime, see also Section 2.3).
The absolute magnitudes of the disc and dipole torques (dotted and 
dashed lines, respectively) are plotted in Figure~\ref{fig:ntime}; 
the short epochs in which the sign of the disc torque is positive 
are plotted with red dots (Figure~\ref{fig:ntime}, left-hand panels).}
The presence of significant (and variable) disc torques causes 
the braking index to vary. More generally, $n$ goes through discontinuities 
caused by transitions between disc-driven spin-up and spin-down phases, between
phases of constant accretion rate and powerlaw decay, and when the location 
of the inner radius of the disc switches between $R_{\rm m}$ and $R_{lc}$. 
In between those rapid transitions, the complex
behaviour of the braking index is mostly determined by the non-linear
coupling between NS spin, inner radius of the disc and accretion
rate.  In particular, for fast pulsars, the dipole torque largely
dominates over the disc torque at early times, while the latter
dominates at later times.  Instead, for slow pulsars, the disc torque
is larger than the dipole component at almost all times.  Both slow
and fast pulsars have a roughly constant braking index during the
first few $10^3$ yrs, corresponding to the epoch when $\dot{M}(R_{\rm
  out}) \approx$ constant and $\dot{M}(R_{\rm in}) \approx \dot{M}_{\rm
  Edd}$. However, the index is exactly $n=3$ for (dipole-dominated)
fast pulsars, while it is $n \approx \Omega/[\Omega-\Omega_K(R_{\rm in})]$ 
for (disc-dominated) slow pulsars (from Eq.(1)).  
The constancy of $n$ in the regime of disc-dominated torque
is expected as long as both $\dot{M}(R_{\rm in})$ and $\Omega$ are roughly constant.

After the first constant phase, there are three types of
discontinuity that can occur on the braking index.
The most dramatic is the transition between spin-down and spin-up phases, 
which formally results in an infinite value for $n$.
These reversals correspond to switches between NS surface accretion and 
propeller phases. In our simulations, they happen only 
at the high end of the initial fallback rate (Fig.~\ref{fig:ntime}, 
left-hand panels); for lower inflow rates, the system remains in the propeller phase 
all the time. Another discontinuity is caused by
the transition of the accretion rate at the inner disc radius from the
early-time Eddington rate to a powerlaw decay. This discontinuity 
in $\ddot{\Omega}$ will
appear only if the disc torque dominates over the dipole torque at the time
when $\dot{M}$ undergoes the transition. 
{Last, as discussed earlier, if, at any point during the pulsar evolution,
$R_{\rm m}$ becomes formally larger than $R_{\rm lc}$, the magnetic pressure
of the pulsar is not able to push the disc further out, and hence the
inner disc boundary tracks $R_{\rm lc}$.  Since the time dependence of
$R_{\rm m}$ (modulated by $\dot{M}(R_{\rm in})$) and that of $R_{\rm lc}$ (dictated
by $\Omega$) are different, these two radii increase at different rates
with time; hence the inner radius of the disc, $R_{\rm in}$, can transition
between $R_{\rm m}$ and $R_{\rm lc}$ (or viceversa) during the coupled
NS/disc evolution.  Any time there is such a transition, the
torque in the right side Eq.(\ref{eq:torque}) changes functional form with time, and hence so does
$\dot{\Omega}$.

We need to remark that our semi-analytical treatment of the torque (Eq. \ref{eq:torque}) naturally
leads to the transitions between the various phases of the coupled
disc/NS evolution to be abrupt. In reality, it is likely that at least
some of these transitions will be smoother. However, we also note
that sharp transitions in the period (and hence discontinuities in
its derivative) are often observed in bynary systems, where the
NS dynamics is known to be driven by the disc (see Bildsten et al.
1997 for a comprehensive study).}

An important feature in the evolution of the braking index
is a period of time during which $n\approx 1$. This 
phase sets in when the disc torque begins to dominate over the 
pulsar torque, and at the same time $\Omega\gg \Omega_K$. Then,
the NS spin is driven by the equation 
$I \dot\Omega\approx -2\dot{M} R^2_m\Omega$,
and the braking index can be written as $n\approx 1+(t/t_*)^{(3/7\alpha-1)}$, where $t_*$
is a constant dependent on a combination of the system parameters. 
For our choice of $\alpha=19/16$, $n\approx 1$ if $t\gg t_*$, 
and the two conditions above are satisfied. 
In our simulations, we find that the (short-lived) $n\approx 1$ phase 
is a tell-tale feature of fast-born pulsars, but is not seen in slow-born 
pulsars, because the latter do not go through a phase with $\Omega\gg\Omega_K$. 

At late times, we find that the disc torque dominates
over the dipole torque both for the slow and the fast pulsars; 
$\Omega$ and $\Omega_K$ track each other, and the torque evolution
is approximately given by $I \dot\Omega\approx 2\dot{M} R^2_m\Omega_K(R_{\rm m})
[1-\Omega/\Omega_K(R_{\rm m})]$. The precise time evolution of the torque depends on 
the index $\alpha$ of the accretion rate. For $\alpha=19/16$,
the time dependence is close to $\propto t^{-1.5}$, like in the case 
of a dipole-dominated torque, and $n\approx 3$. Larger values of $\alpha$ 
would yield somewhat smaller values of $n$ at late times.

\begin{figure}
\centering
\vspace{-0.1in}
\includegraphics[width=8.4cm]{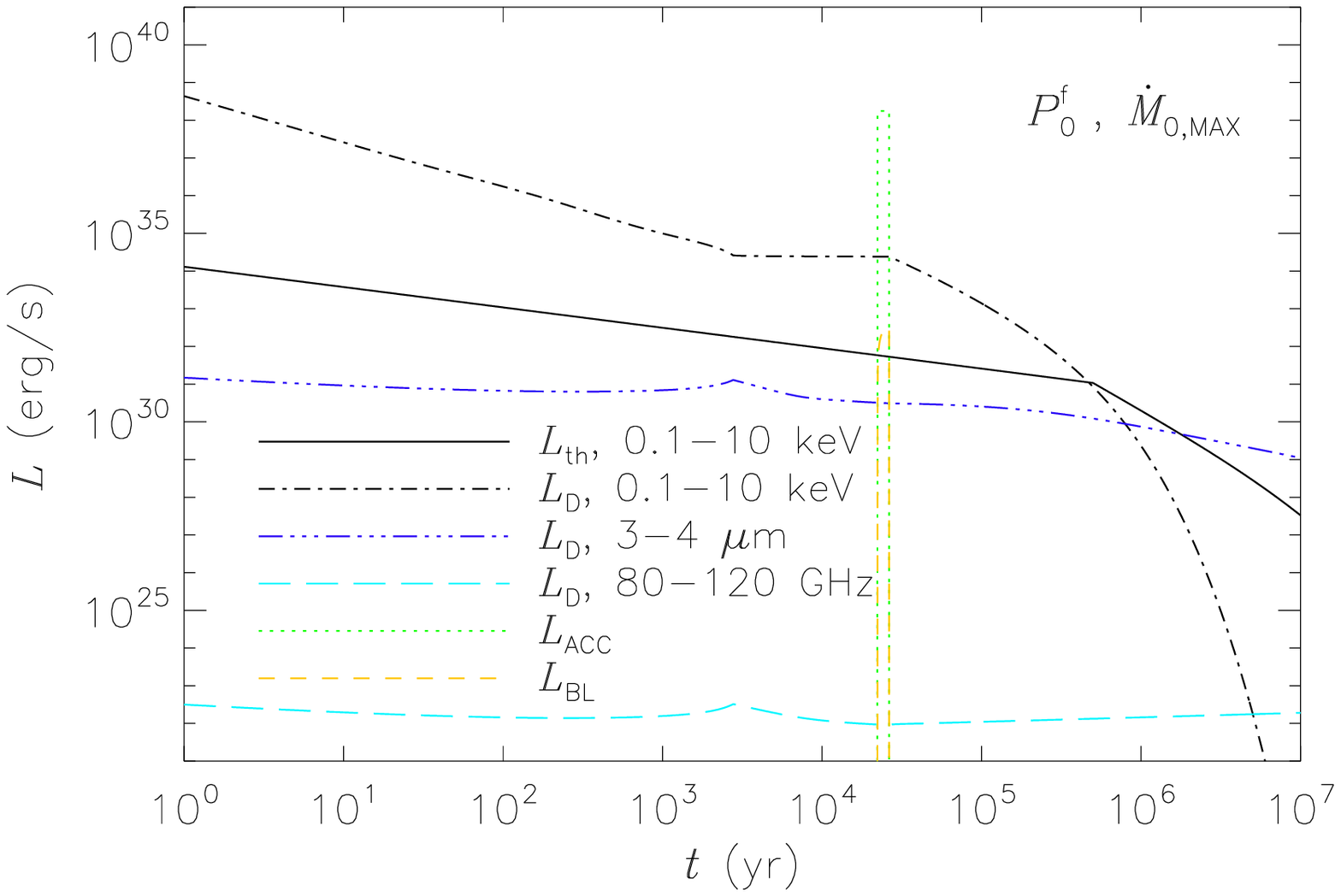}\\[-13pt]
\includegraphics[width=8.4cm]{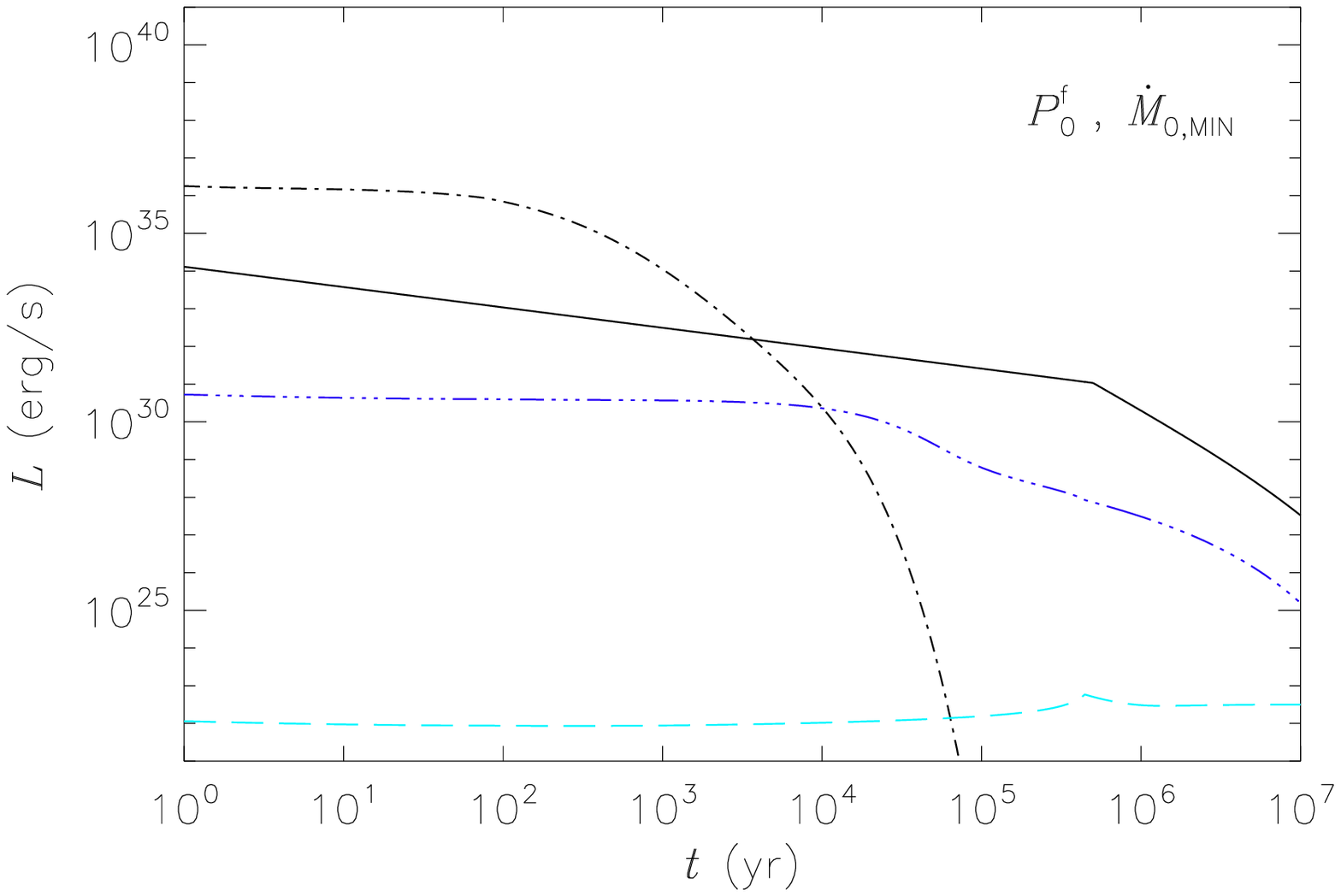}\\[-13pt]
\includegraphics[width=8.4cm]{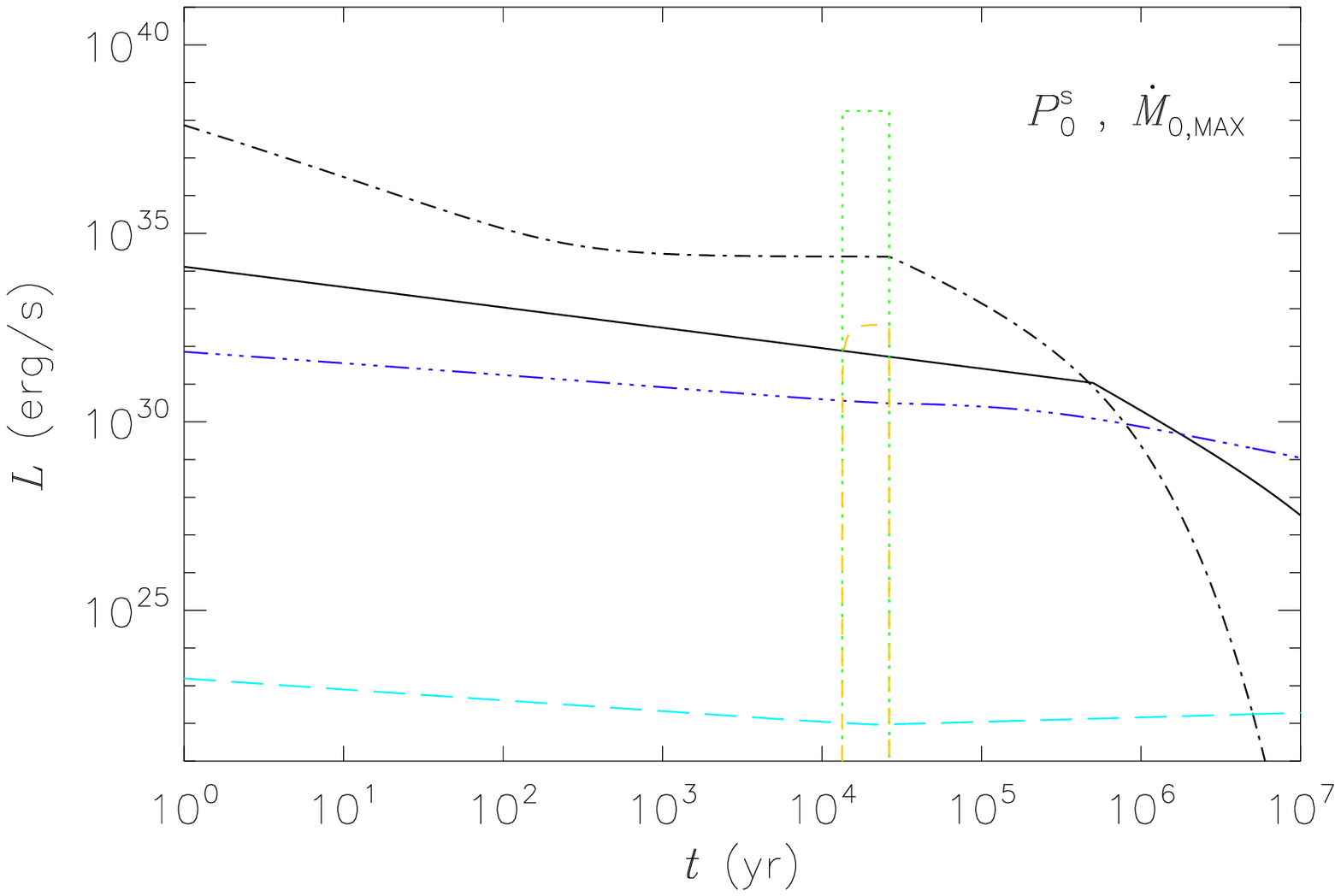}\\[-13pt]
\includegraphics[width=8.4cm]{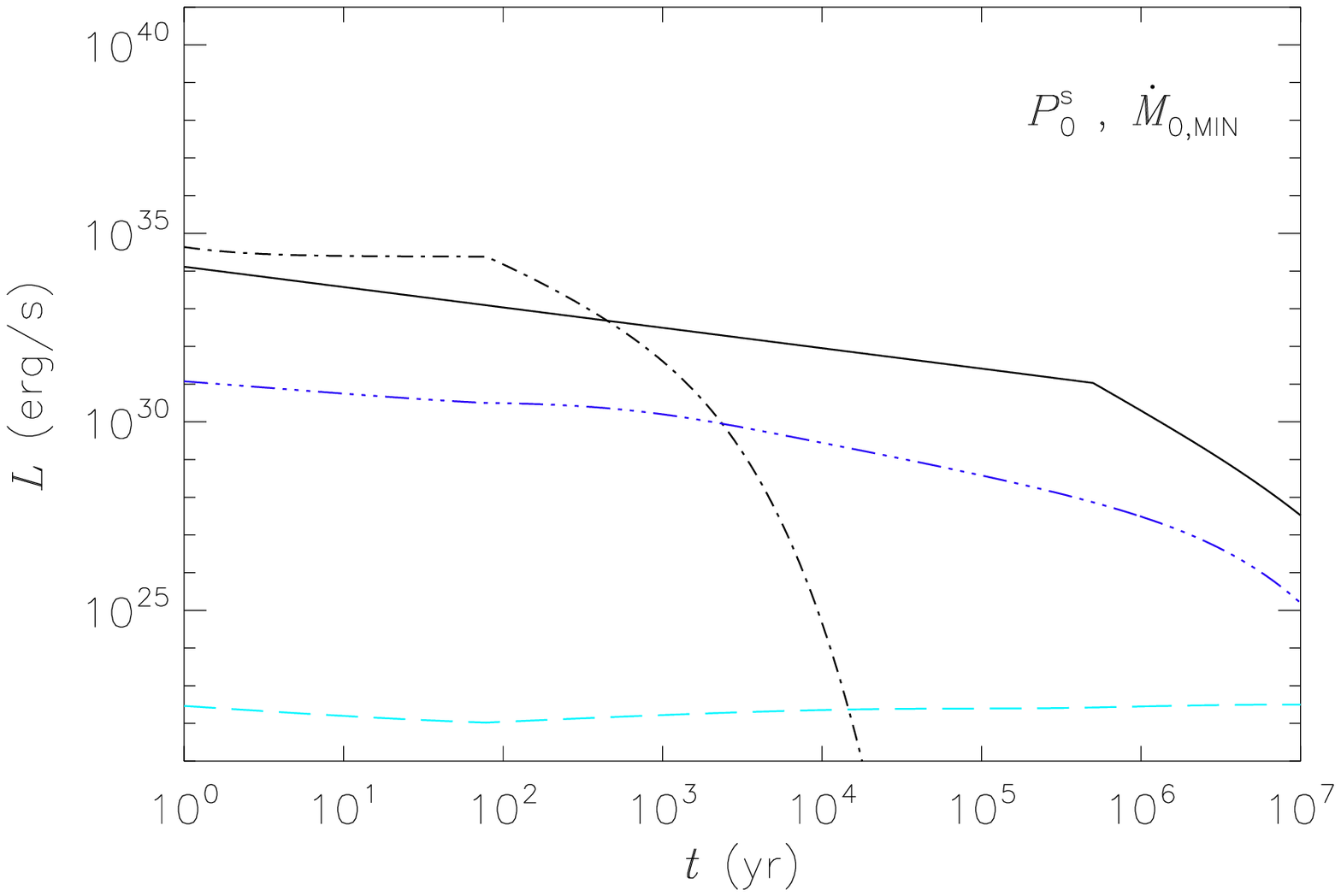}\\[-13pt]
\caption{Time evolution of the various components of the NS+disc luminosity, for 
fast-born ($P_0^f=5$~ms) and slow-born ($P_0^s=300$~ms) pulsars, accreting
at $t=0$ with either a high ($\dot{M}_{0, {\rm max}}=10^{28}$g~s$^{-1}$), or a low  
($\dot{M}_{0, {\rm min}}=10^{25}$g~s$^{-1}$) inflow rate. $L_{\rm th}$ indicates the
thermal surface emission of the NS.} 
\label{fig:Ltime}
\end{figure}

\subsection{Luminosity Evolution}

The luminosity of an NS with a fallback disc is given by the sum of
two terms: {a thermal component due to the hot surface of the NS}, and 
another thermal component (multi-blackbody) due to radiatively-efficient 
viscous dissipation in the accretion flow.
The ultimate source of power is {internal heat} for the first term, 
and the gravitational potential drop of the fallback matter, 
for the second term. In turn, the accretion power can be split into 
separate contributions from the accretion disc, from the boundary layer, 
and from the NS surface. {Furthermore, when the disc does not penetrate
inside the light cylinder (i.e. when $R_{\rm in}=R_{\rm lc}$), the 
pulsar mechanism is likely to operate and then magnetospheric emission, powered
by the pulsar spin down, is also present.}

More in detail, the various emission components are as follows.

\medskip
\noindent (i) {\em   NS emission}. 
{The main component, always present with
or without a disc, is the surface emission of the NS, due to its
internal cooling.  We model the resulting X-ray luminosity using the
standard cooling model of Blaschke et al. (2004), for a 1.4$M_\odot$
NS, with radius of 10~km. Furthermore, as discussed above, there can
be periods during the coupled NS/disc evolution when the pulsar
mechanism is able to operate, converting a fraction of the spin power
of the NS into non-thermal X-ray radiation.} Observationally, a
correlation between the spin power $\dot{E}_{\rm rot}$ and the 2-10~keV
luminosity has been investigated in numerous studies (e.g. Saito et al.
1997; Possenti et al. 2002; Kargaltsev \& Pavolov 2008). In particular,
Possenti et al. (2002) derive the best-fitting relation
\begin{equation}
\log L_{X,[2-10]}=1.34\log\dot{E}_{\rm rot} - 15.34\,,
\label{eq:LxPul}
\end{equation}
with $L_X$ and $\dot{E}_{\rm rot}=I\Omega\dot{\Omega}$ in CGS units. 
{When present, such a component would manifest itself as a power-law spectrum
of photon index $\Gamma\approx 2$.}

{In this work, since we are specifically interested in the properties
of fallback discs, and their effects on NSs, we compute in detail the
various components of the emission due to the disc itself (viscous
dissipation, irradiation) and its interaction with the NS (boundary
layer, accretion). However, since the surface emission of the NS
cannot be hidden, we also include this contribution for reference and
comparison to the components above. For the magnetospheric component,
on the other hand, we will limit ourselves to a discussion, since the
presence of this component depends on the location of $R_{\rm in}$
at a given time, and its magnitude is determined by the
timing parameters of the NS at that time.}

\medskip
\noindent (ii) {\em  Surface accretion luminosity}. This term is given by the 
potential and kinetic energy
released by matter falling from the magnetospheric radius to the
surface of the NS; hence, in the Newtonian approximation
it is:
\begin{equation}
L_{\rm acc}= GM\dot{M}\left(\frac{1}{R_{NS}}-
\frac{1}{R_{\rm in}}\right)+\frac{1}{2}\Omega^2\left(R_{\rm in}^2-R_{NS}^2\right)\;.
\label{Lac}
\end{equation}
For the characteristic radius of a NS ($\approx 10$ km), the effective 
blackbody temperature of the surface accretion luminosity 
spans the soft X-ray band, between $\sim 0.1$ keV and $\sim 2$ keV, 
for a large range of luminosities, from those characteristic 
of Bondi accretion from the instellar medium \citep{zam95} 
up to the Eddington luminosity. 
It is important to notice that this component is present only
when $R_{\rm in}=R_{\rm m}<R_{\rm co}$, where $R_{\rm co}$ is the corotation radius,
i.e. the radius at which the Keplerian rotation of the material
is equal to the rotation rate of the star $\Omega$ (condition equivalent to
 $\Omega<\Omega_{\rm K}(R_{\rm m})$).
Otherwise, the pulsar is found in the propeller
phase, accretion is inhibited, and the corresponding luminosity
contribution can be considered negligible.\\

\medskip
\noindent (iii) {\em  Boundary layer luminosity}. 
This contribution is due to the release of energy in the boundary layer 
between the Keplerian disc and the magnetosphere. 
We estimate the boundary layer luminosity with the following argument, 
following Perna et al. (2006).
We adopt the ``elasticity parameter'' $\beta$, which is a measure of
how efficiently the kinetic energy of the NS is converted
into kinetic energy of ejected matter through the magnetosphere-disc interaction
(Eksi et al. 2005). In the limit of a completely anelastic propeller 
($\beta=0$), the magnetosphere forces matter to corotate with it
during both the accretion and the propeller regime. On the other hand, 
in the limit of a completely elastic propeller
($\beta=1$), matter is slowed down in
the boundary layer; this term is present only during the
propeller phase. 

For a generic value for the elasticity parameter, the
luminosity of the boundary layer can be written as:
\begin{equation}
L_{\rm BL}=\left\{ \begin{array}{ll}
 \displaystyle\frac{\dot{M}}{4\pi}\big[R_{\rm m}^2(\Omega_{\rm K}^2-\Omega^2)\big]
 & {\rm for}\;\; R_{\rm m}<R_{\rm co} \\[10pt]
 \displaystyle\frac{\dot{M}}{4\pi}(1-\beta)\big[R_{\rm m}^2(\Omega^2-\Omega_{\rm K}^2)\big]
 & {\rm for}\;\;R_{\rm m}\geq R_{\rm co}\;.
\end{array} \right.
\label{lbl}
\end{equation}
Given the uncertainties in the parameter $\beta$, here we adopt
$\beta=1$, which yields a lower limit to the attainable boundary layer
luminosity during the lifetime of the NS.

\medskip
\noindent (iv) {\em Disc luminosity}.
This contribution is clearly present in both the accretion and 
the propeller regime. 
For an optically thick, geometrically thin disc model \citep{sha73,fra02}, 
the luminosity in the $[\nu_1,\nu_2]$ frequency 
band is given by an integration over radial annuli, 
\begin{equation}
L_{[\nu_1-\nu_2]}= 2\pi\int_{\nu_1}^{\nu_2} d\nu\frac{h\nu^3}{c^2}\int_{R_{\rm in}}^{R_{\rm out}}\frac{R\, dR}
{e^{h\nu/k T(R)}-1}\;,
\label{eq:bb}
\end{equation}
each producing a blackbody spectrum at a temperature 
\begin{equation}
T(R) = \left(\frac{3GM_{\rm NS}\dot{M}(R)}{8\pi R^3\sigma}\right)^{1/4}\left[
1-\left(\frac{R_{\rm in}}{R}\right)^{1/2}\right]^{1/4}\;.
\label{eq:temp}
\end{equation}
Besides emission due to internal energy dissipation, 
the disc can brighten up also as a result of reradiation of 
X-rays impinging from the central object. We use Vrtilek et al.'s (1990) 
analytical expression for the effective temperature of an
irradiated disc, under the assumption that the disc height
$h\propto r^n$ and irradiation is the dominant form of heating.
If particles in the disc are in Keplerian motion,
then $n=9/7$, and the temperature profile is given by
\begin{eqnarray}
&&T_X(r) = \left[f\frac{\sqrt{k_B/\mu m_{\rm H}} L_X \omega}
{14\pi\sigma GM}\right]^{2/7}  \nonumber \\
& \simeq & 2.32 \times 10^4  
\left(\frac{f}{0.5}\right)^{2/7}\left(\frac{L_X}{L_{\rm Edd}}\right)^{2/7}
\left(\frac{R_\odot}{r}\right)^{3/7} K,
\label{eq:t2}
\end{eqnarray} where $L_X$ is the $X$-ray luminosity of the central source, and
$f$ parameterizes the uncertainty in the disc structure and albedo.  
To a first approximation, the total luminosity in the $[\nu_1,\,\nu_2]$ band 
is the sum of the viscous dissipation and reradiation components; 
the former dominates at small radii, the latter dominates in the outer disc.

Recall that we are not assuming a constant $\dot{M}$ throughout the disc. 
As we discussed in Section 2.1, at earlier times $\dot{M}(R) \sim R$ 
because of wind losses.
Therefore, during the early super-critical accretion phase, 
the temperature profile in the inner disc is $T(R) \sim R^{-1/2}$, flatter 
than the standard disc-blackbody profile, which scales as $R^{-3/4}$. 
Our choice of accepting an Eddington-limited accretion rate 
at the innermost disc radius and a much higher mass inflow rate 
at the outer boundary implies that the integrated disc luminosity 
may exceed $L_{\rm Edd}$ at early times. However, it is easy 
to see that for $\dot{M}(R) \sim R$, this excess is only a logarithmic 
function of $\dot{M}(R_{\rm out})/\dot{M}_{\rm Edd}$. This is a restatement 
of the well-known result that 
$L_{\rm disc} \approx [1 + \ln(\dot{M}/\dot{M}_{\rm Edd})]$
during super-Eddington accretion \citep{sha73,fra02}.
In our case, the highest possible inflow rate at the outer radius 
considered in our model, $\dot{M}(R_{\rm out})/\dot{M}_{\rm Edd} \sim 10^{10}$,
would lead to a bolometric disc luminosity $\sim 20\,L_{\rm Edd}$, 
at least for the initial transient phase (or less, if $R_{\rm m} \gg 10$ km). 
We accept this possibility of transient super-Eddington luminosity 
as we are modelling a situation characterized by massive outflows 
rather than a steady-state solution. A more complex modelling of 
the accretion flow in that situation with a slim-disc rather than 
a standard-disc plus outflow approximation is beyond the scope of this work.

We begin by exploring the relative magnitude of the various components
of the luminosity as a function of NS age, for a few
representative combinations of NS and disc parameters. In
particular, for the same choice of initial NS spin and disc
accretion rate as in Fig.~\ref{fig:ntime}, we consider the
time-evolution of the
surface accretion, boundary layer, and the disc luminosity in the
$0.1$--$10$~keV, the 3-4$\mu$m, and in the 80-120 GHz bands.  
In the IR
and the mm band we compute only the disc luminosity, since the pulsar
contribution in these bands is not well known.  
For comparison, we also show the timing evolution of the NS thermal emission in 0.1-10~keV.
We assume an outer
disc radius\footnote{In reality, the outer radius of the disc is
  expected to be an increasing function of time. Modeling this in
  detail is however beyond the scope of our paper, since the magnitude of the disc
   torque depends only on the location of the inner boundary
  of the disc. The X-ray and IR luminosity of the disc is also not very
  sensitive to the outer radius (unless very small), since it is
  produced relatively close to $R_{\rm m}$, while the mm luminosity is more
  sensitive to the exact value of $R_{\rm out}$, see e.g. Posselt et
  al. (2010).} $R_{\rm out}=1000 \,R_{\rm NS}$, and a face-on
geometry. Hence the predicted luminosities represent upper limits for
discs with generic inclinations with respect to the observer line of
sight.

\begin{figure*}
\centering
\includegraphics[width=8.2cm]{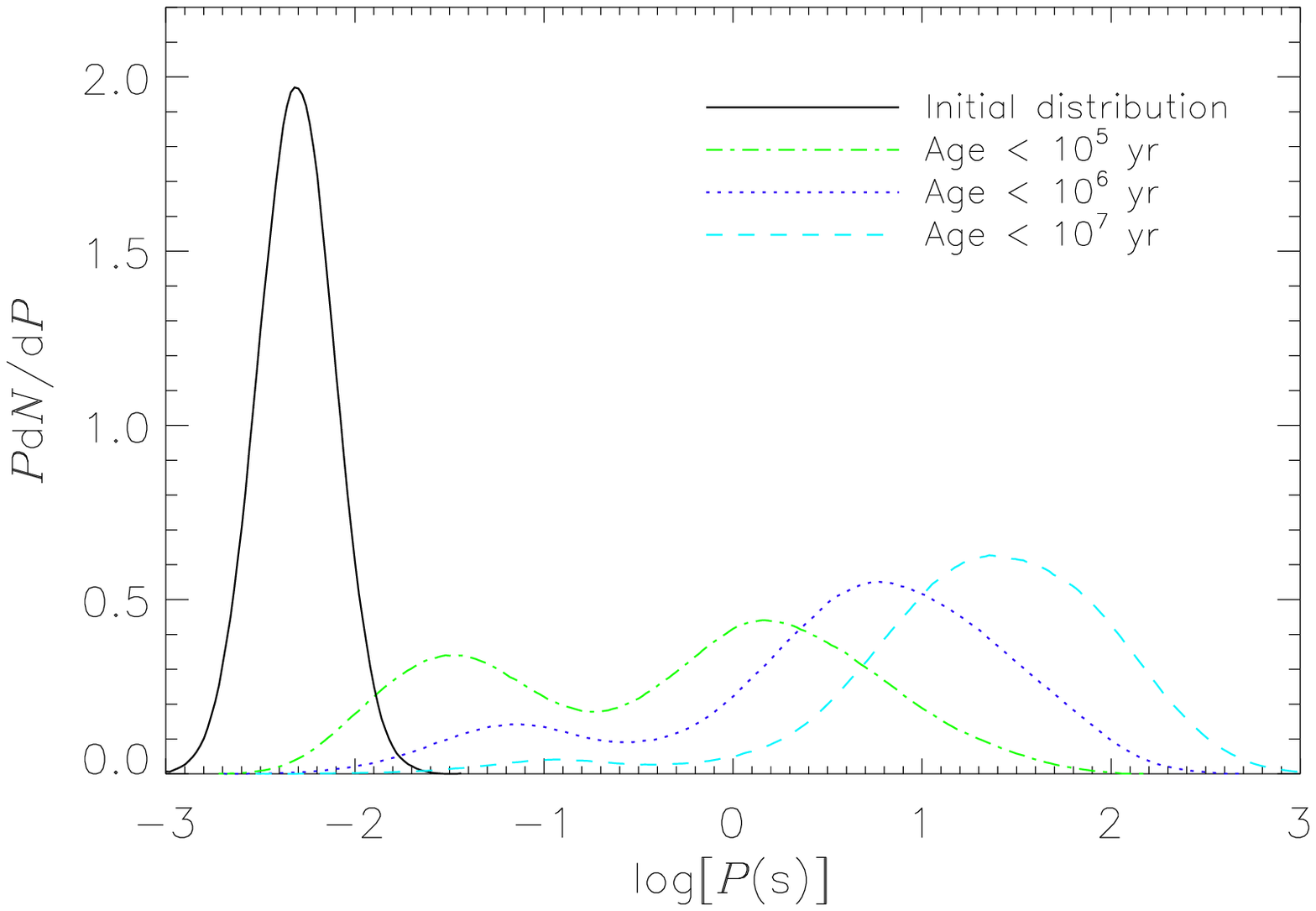}
\includegraphics[width=8.2cm]{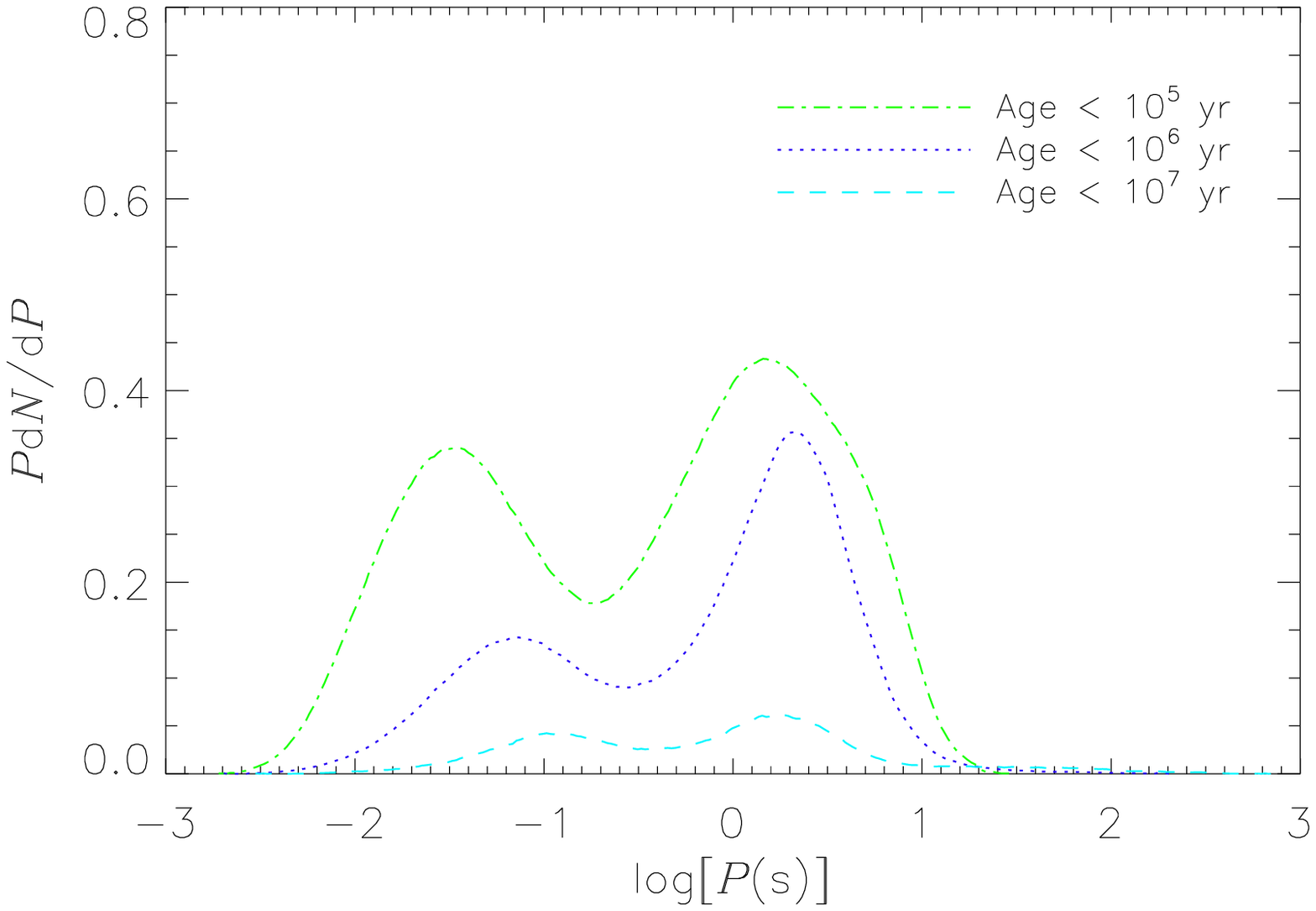}\\
\includegraphics[width=8.2cm]{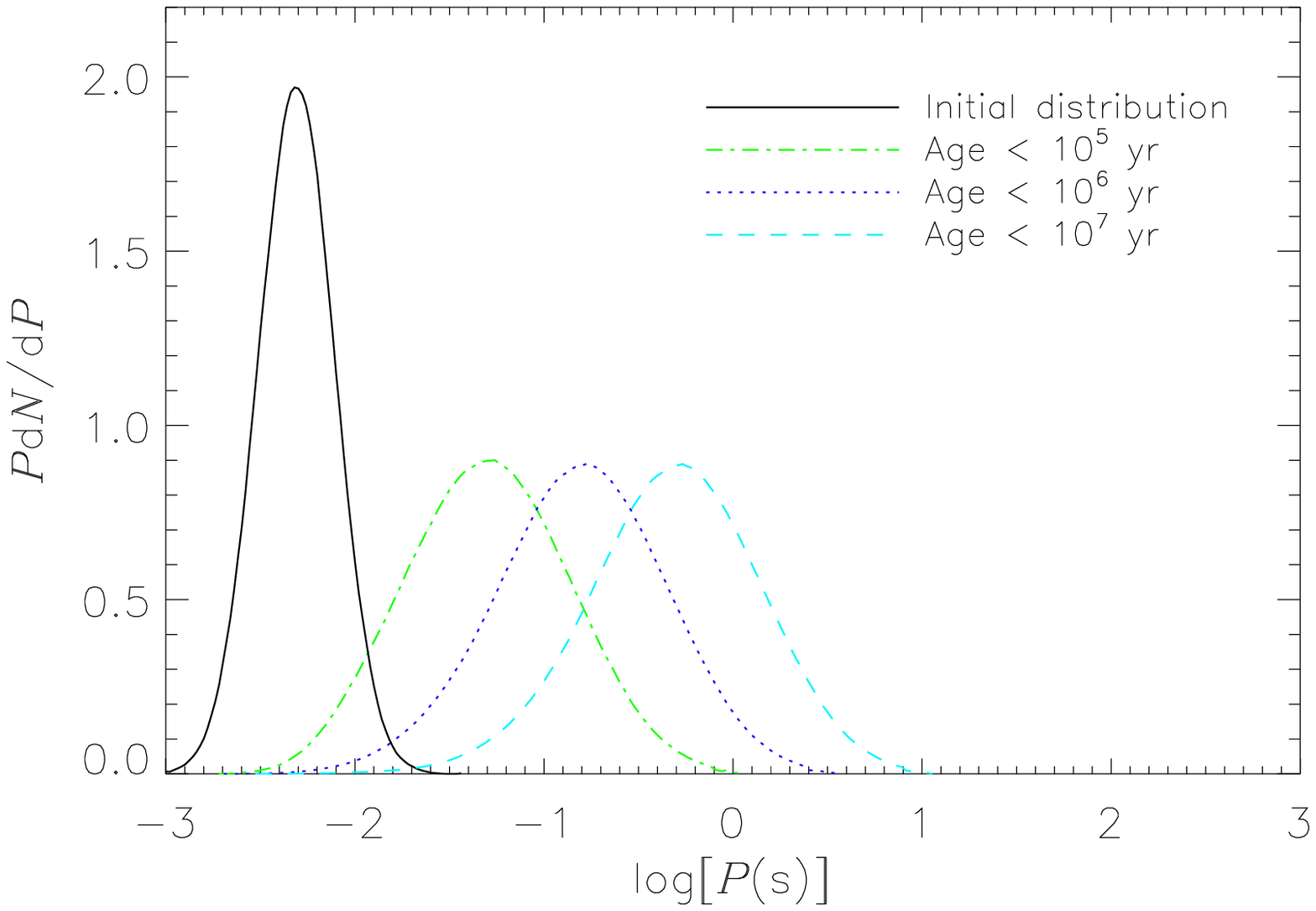}
\includegraphics[width=8.2cm]{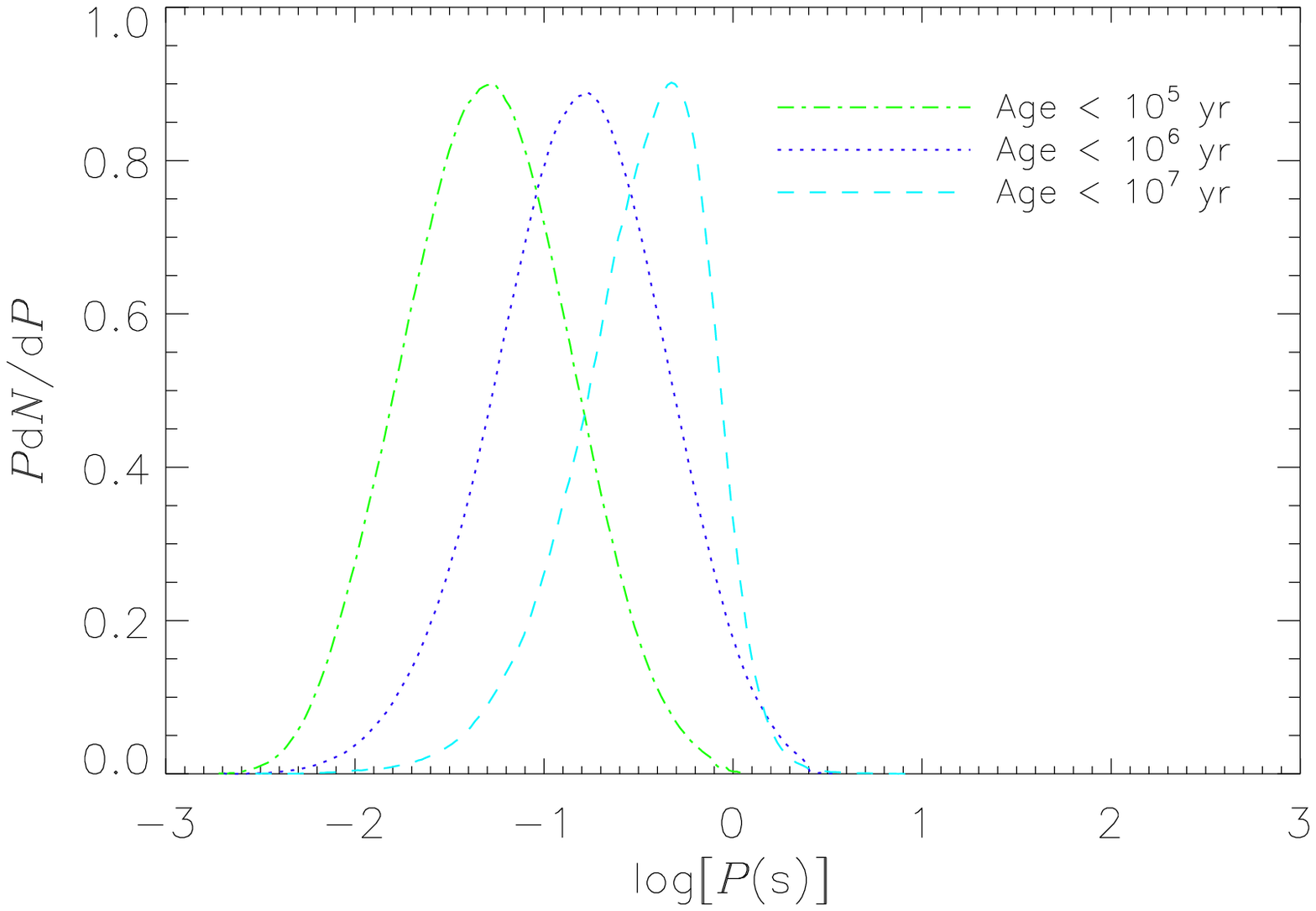}
\caption{Left: the evolution of a representative,
  short-period NS spin distribution at birth (solid line) for the
  case in which the NSs evolve by dipole radiation alone (bottom
    panel), and for the case in which they evolve as the result of
  dipole losses together with magnetospheric interaction with a
  fallback disc (upper panel). Right: same as in the left-hand panels,
but including only the pulsars above the death line.}
\label{fig:Pfast}
\end{figure*}

Fig.~\ref{fig:Ltime} shows the evolutionary path of the various
components of the luminosity, for the four combinations \{$\dot{M}_{0,
  {\rm max}},P_0^f$\}, \{$\dot{M}_{0, {\rm max}},P_0^s$\},
\{$\dot{M}_{0, {\rm min}},P_0^f$\} and \{$\dot{M}_{0, {\rm
    min}},P_0^s$\}, where $\dot{M}_{0, {\rm max}} = 10^{28}$g~s$^{-1}$
and $\dot{M}_{0, {\rm min}} = 10^{25}$g~s$^{-1}$.  We suggest the
following physical interpretation of these results.  At the low end of
the initial fallback rate, the NS will never be found in an accretion
phase; hence, both the surface accretion and the boundary layer
luminosities will always be negligible.  In the {0.1-10~keV band,
  it is interesting to note that the disc luminosity exceeds that from
  the NS surface under all choices of parameters, for an age $t\la
  10^3$~yr.  The higher the accretion rate, the longer is the time
  over which the disc luminosity exceeds that of the NS (compare
  panels labeled with $\dot{M}_{0,{\rm MAX}}$ with those labeled
  $\dot{M}_{0,{\rm MIN}}$ in Fig.\ref{fig:Ltime}).  On the other hand,
  the magnetospheric component of the luminosity due to rotational
  energy losses, when the pulsar mechanism is active, would produce
  very high luminosities at early times for millisecond pulsars (see e.g. Perna \& Stella
  2004). However, spectral modeling can in principle  disentangle the
  thermal disc component from the powerlaw emission from the
  magnetosphere.}

When the initial fallback rate is very high, we have already noted 
(Section 2.2) that, at some epochs, $\Omega_{\rm K}>\Omega$, and the NS is
temporarily spun up while accreting. During these transient bright phases, the
X-ray brightness of the system is dominated by the accretion luminosity
onto the surface of the NS.  
The exact duration of this transient bright phase, and the time at which it sets
in, are a function of $P$, $\dot{M}$ and $B$; a more detailed discussion 
is beyond the scope of this work. The physical significance of these 
accretion-powered phases is that, if fallback discs
around pulsars are ubiquitous, and they set in at high accretion rates, 
then they would give rise to {\em a new class of X-ray transients}.
{Furthermore, we note that a sudden increase in brightness could also occur if,
at some point during the evolution,  the
inner radius switches from $R_{\rm in}<R_{\rm lc}$ to $R_{\rm in}=R_{\rm lc}$, 
at which point the pulsar mechanism can turn on again.}

As already discussed in previous work on fallback discs, the disc is
expected to be brighter at wavelengths longer than the X-rays. This is
clearly seen by the mid-IR disc luminosity (computed in the 3-4~$\mu$m
band) also diplayed in Fig.~\ref{fig:Ltime}; the mid-IR luminosity is,
especially at later times (when the disc is cooler), substantially
larger than the 2-10~keV luminosity (see \citealt{per00a,per00b} for
detailed spectra of fallback discs).  Searches for fallback discs in
the optical and IR have been numerous, but they have mostly yielded
upper limits on the emission.  Mignani et al. (2007a,b) derived an
upper limit on the K luminosity of the pulsar PSR~J1119-6127 of
$6.6\times 10^{30}$~erg~s$^{-1}$.  A dedicated search for fallback
discs in four supernova remnants was performed by Wang et
al. (2007a). Their IR limits are on the order of a few $\times$
$10^{29}$~erg~s$^{-1}$ for objects of a few kyr of age.  Taken at face
value, these limits, once compared with the results of
Fig.~\ref{fig:Ltime}, would imply that fallback discs, if ubiquitous
around young NSs, are unlikely to be accreting at very high rates
and/or be very large (recall that here we have assumed $R_{\rm
  out}=1000\,R_{\rm NS}$).  Detection of a fallback disc around an
isolated NS was reported by Wang et al. (2006). The mid-IR luminosity
was $8.5\times 10^{31}$~erg~s$^{-1}$ (for a distance of 5~kpc). This
value is compatible with our predictions.  More generally, IR emission
has been detected in a number of isolated NSs (see e.g. Mignani 2011
for a summary).  However, lack of a detailed spectral characterization
makes it typically difficult to distinguish between magnetospheric and
disc emission.

Longer wavelength observations (in the submillimeter) have been
reported by Posselt et al. (2010) for the isolated NS 
RX~J1856.5-3754.  Using the bolometer array on the APEX telescope,
they determined a flux limit of 5~mJy at 345~GHz. At the estimated
distance of 167~pc, this yields a limit of $5.8\times
10^{28}$~erg~s$^{-1}$ on the luminosity.  This value is unconstraining
for the discs considered here. And in fact, Posselt et al. (2010)
found that it could only set limits on a hypothetical very large outer
radius of the disc.  The best prospect for detecting the presence of
discs around young NSs remains in the optical/IR wavelength range.

\begin{figure*}
\centering
\includegraphics[width=8.2cm]{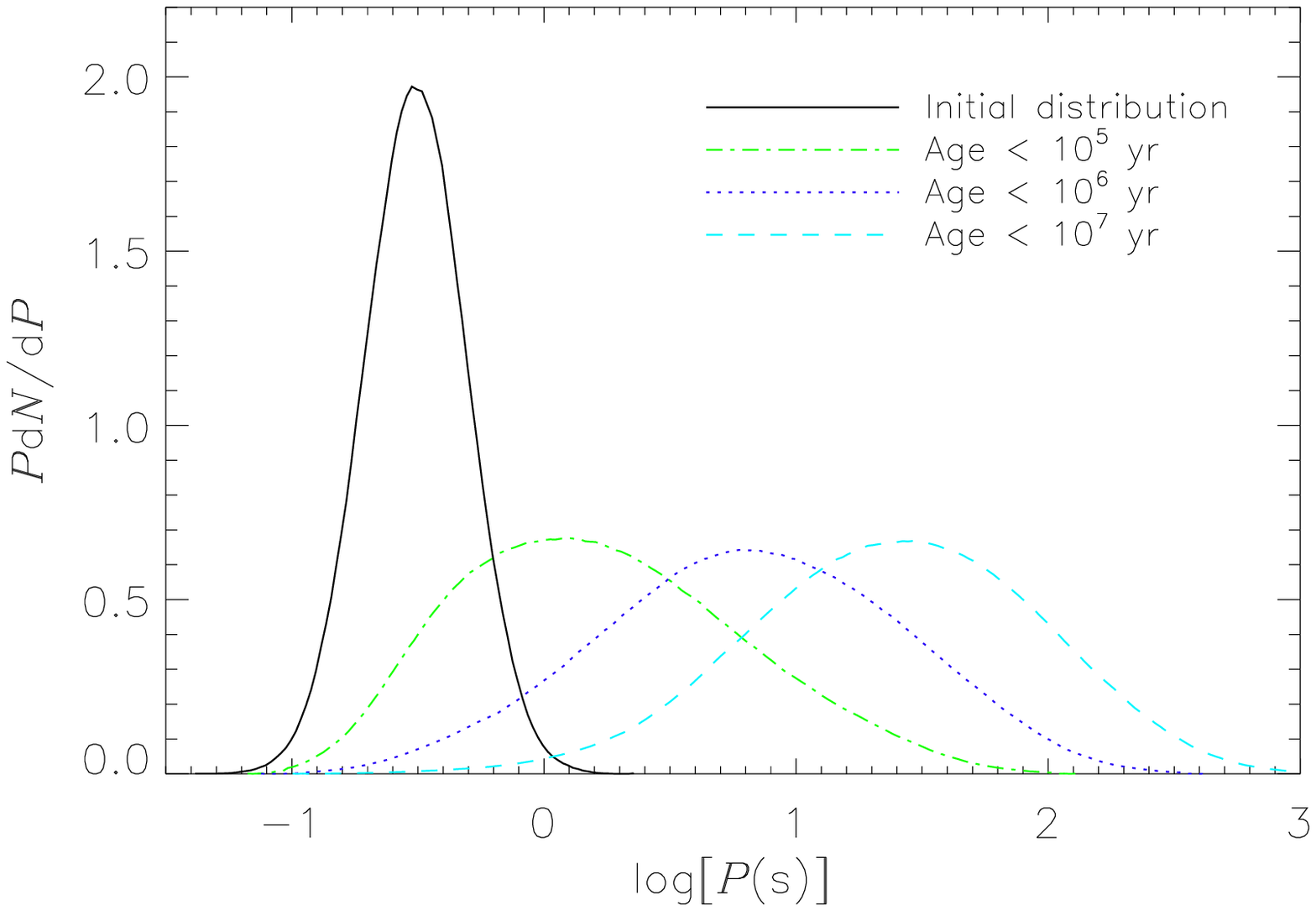}
\includegraphics[width=8.2cm]{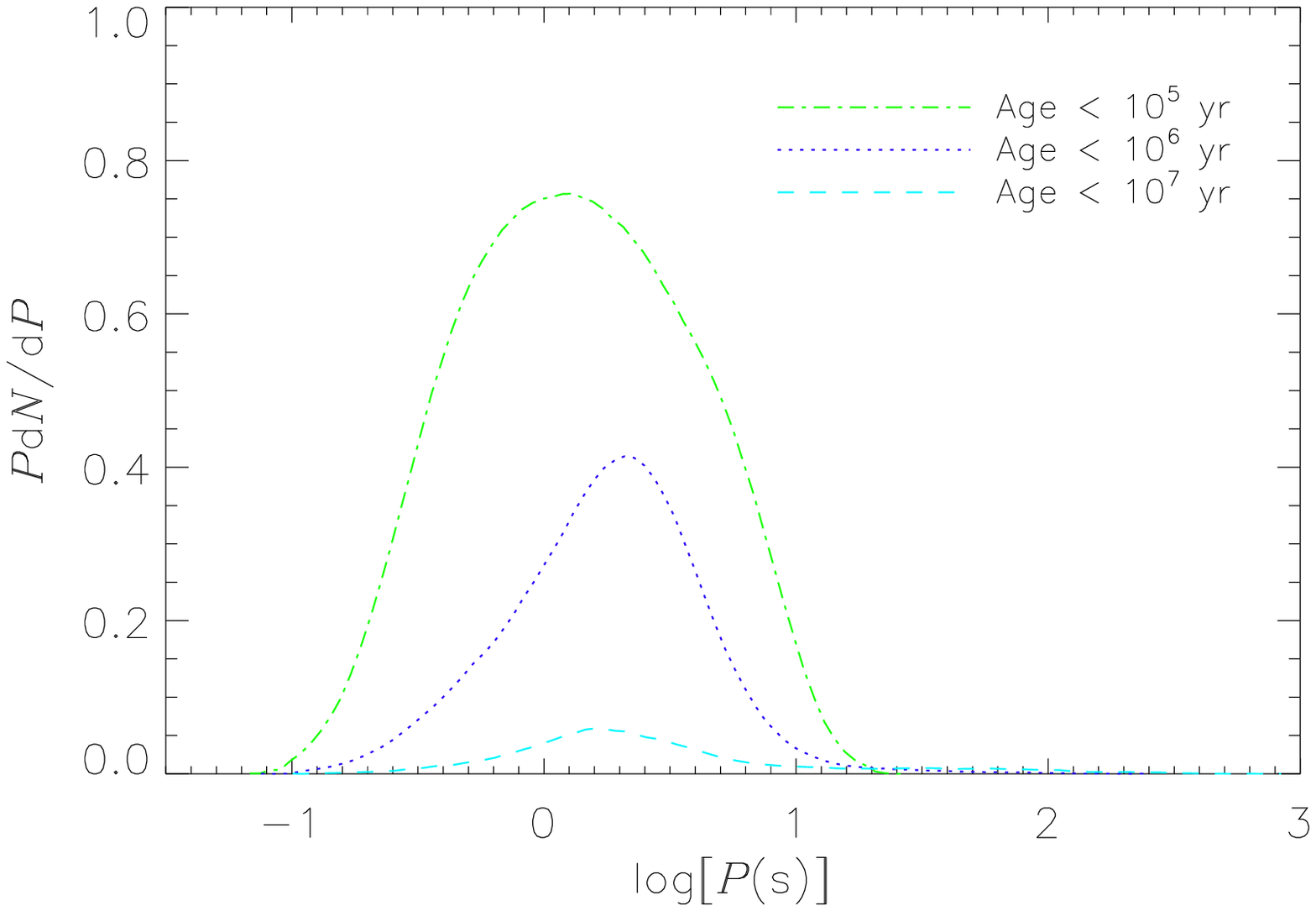}\\
\includegraphics[width=8.2cm]{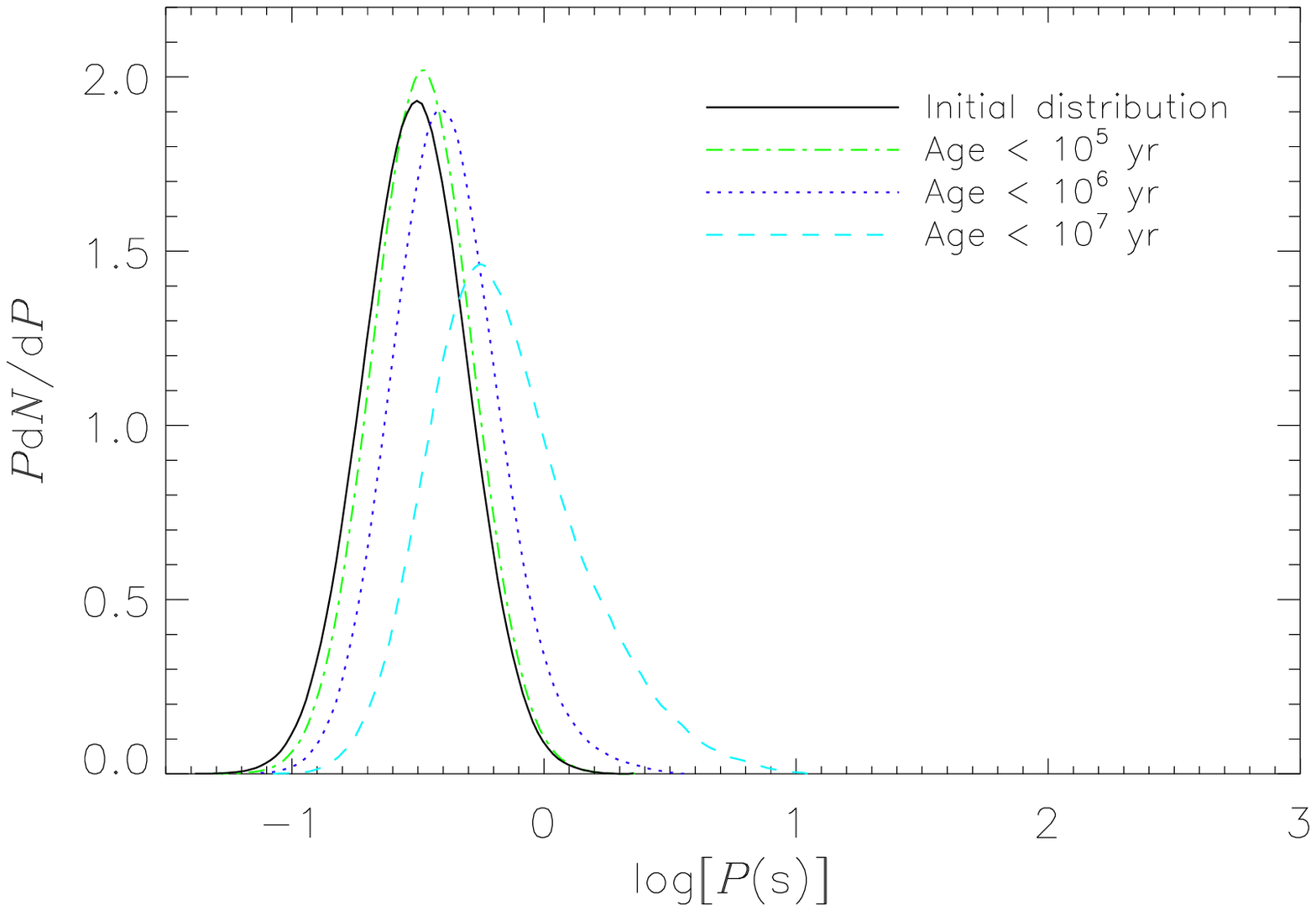}
\includegraphics[width=8.2cm]{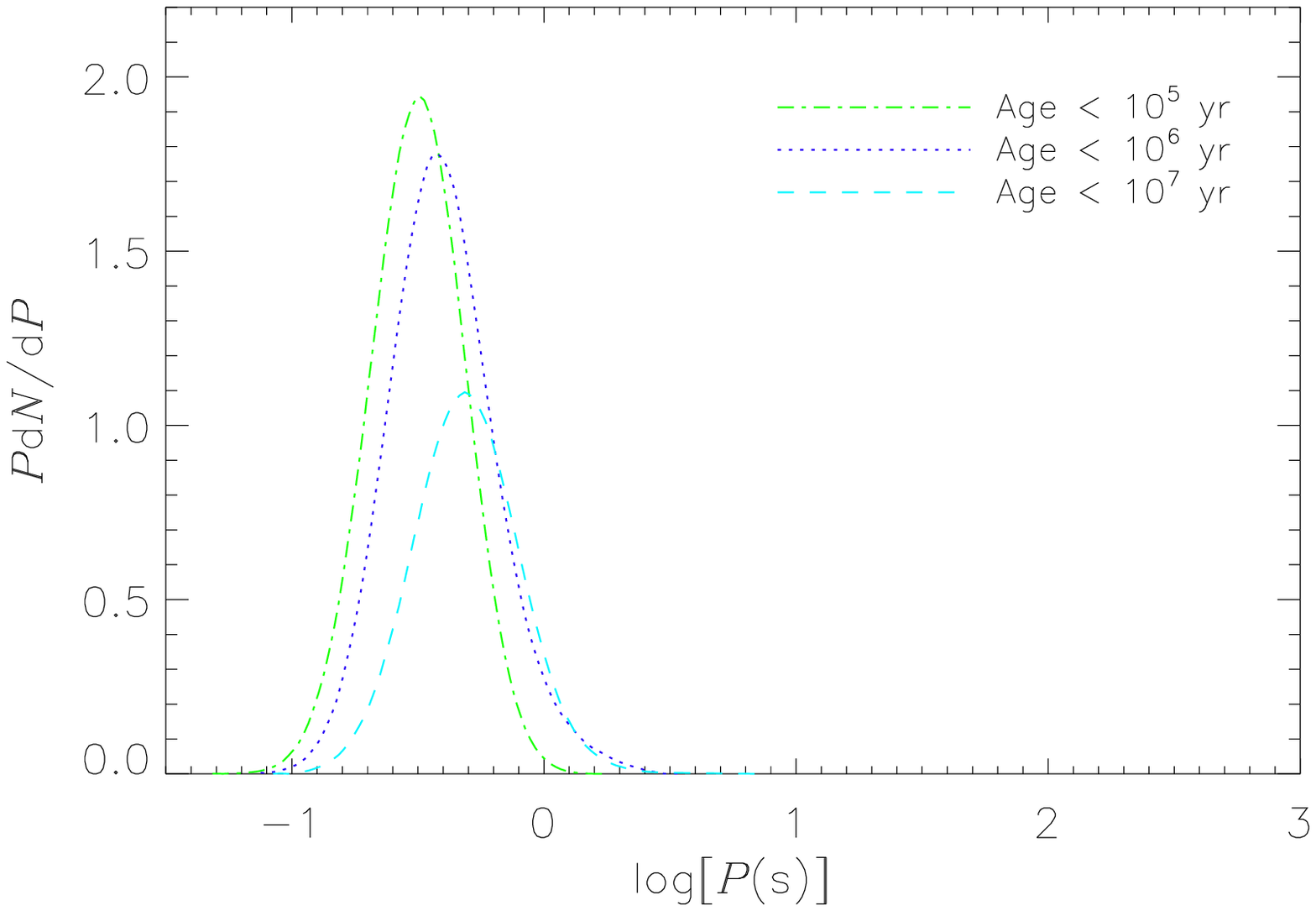}
\caption{Same as in Fig.~\ref{fig:Pfast}, but for the long-period spin
distribution at birth.}
\label{fig:Pslow}
\end{figure*}

\section{Population properties}

\subsection{Timing Properties}

We proceed now to simulate how the ubiquitous presence of fallback discs affect the
statistical timing properties of the NS population, for a certain
birth spin period and magnetic field distribution. 
The magnetic field $B$, and the initial disc inflow rate $\dot{M_0}$ are 
randomly generated according to their respective distributions discussed in
Section 2.2.  For the initial spin, we consider two distributions: 
one centred at $P_0 = 5$ ms (representative of fast-born pulsars) 
and one at $P_0 = 300$ ms (slow-born pulsars);
we assume $\sigma_{\log P_0} = 0.2$ for both distributions.
Finally, for plotting purposes, we normalize the results to a pulsar birth rate 
$\dot{N} = 1/100$ yr$^{-1}$. 

In order to simulate the distribution of {\it observable} radio pulsars, 
we need to take into account 
that the radio emission fades with age. Here, we adopt 
the death probability band empirically defined by \citet{arz02}, 
centered around the line $\dot{P}_{15}/P^3 -10^{\rm DL_0}=0$, 
where $P$ is in the seconds, $\dot{P}_{15}\equiv \dot{P}/10^{-15}$~s/s, 
and DL$_0=0.5$.
The radio emission can also be disrupted or modified when the disc
penetrates the magnetosphere inside the light cylinder (e.g. Shannon
\& Cordes 2009 for an extensive discussion). This does not affect the
long-term evolution of the period and timing properties computed
here. However, if disc penetration inside the magnetosphere were to
dim or completely shut down radio emission, then a fraction of pulsars
in the sample may not be visible. Previous investigations (e.g. Alpar
et al. 2001; Menou et al 2001a) assumed the inner edge of the disc to
be always located at the light cylinder, throughout the entire
evolution of the pulsar, so that radio emission would always be
ensured. However, when the accretion rate is very high, and the ram
pressure of the material is larger than the magnetic pressure of the
pulsar, there is no compelling reason for imposing that the inner edge
of the disc does not penetrate inside the light cylinder. Hence, in this
investigation, we let the disc torque operate at its physically
expected location, that is the smallest between $R_{\rm m}$ and $R_{lc}$,
with the understanding that, if $R_{\rm m}<R_{lc}$ the radio signal might be
disrupted.
With this caveat in mind, in the rest of the paper we conform to the
standard language in the literature, and refer to 'dead' pulsars only
as to the fraction of pulsars that have passed the death line.  
For a given distribution of initial parameters, 
we calculate period distributions for both the full NS population 
(whether observable or not) and for the sub-population 
of observable radio pulsars (before the death-line cutoff).

\begin{figure*}
\centering
\includegraphics[width=8.2cm]{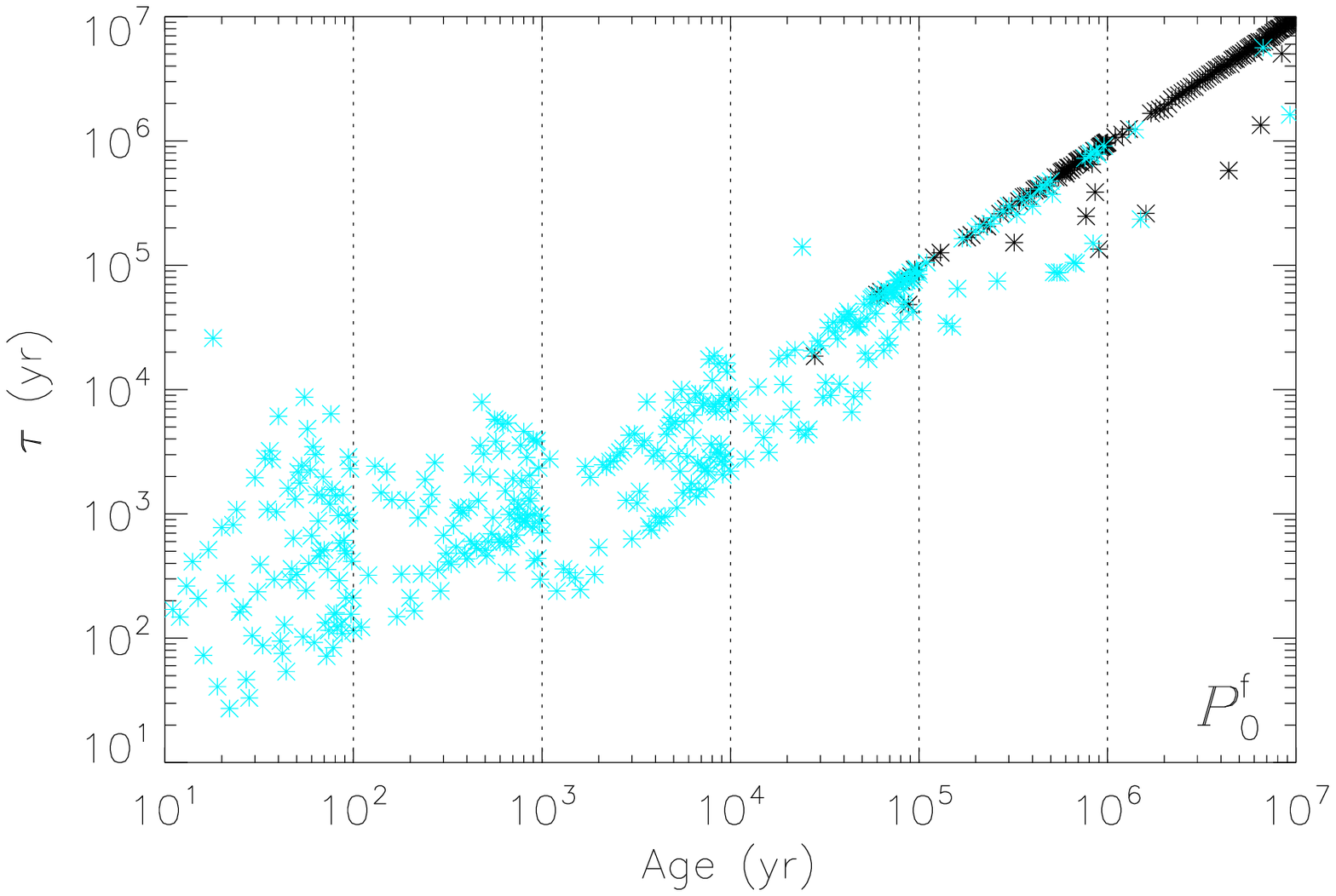}
\includegraphics[width=8.2cm]{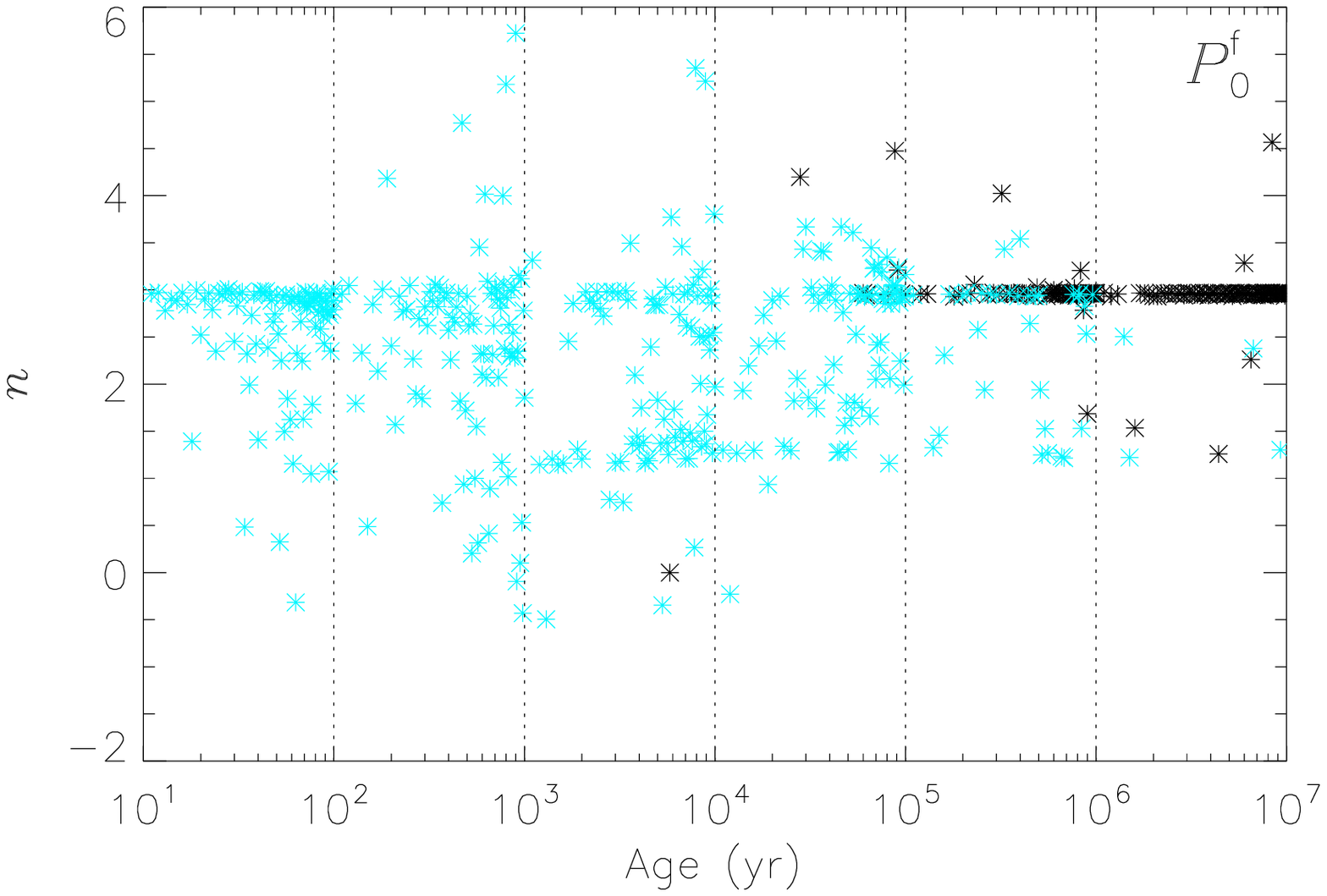}\\
\includegraphics[width=8.2cm]{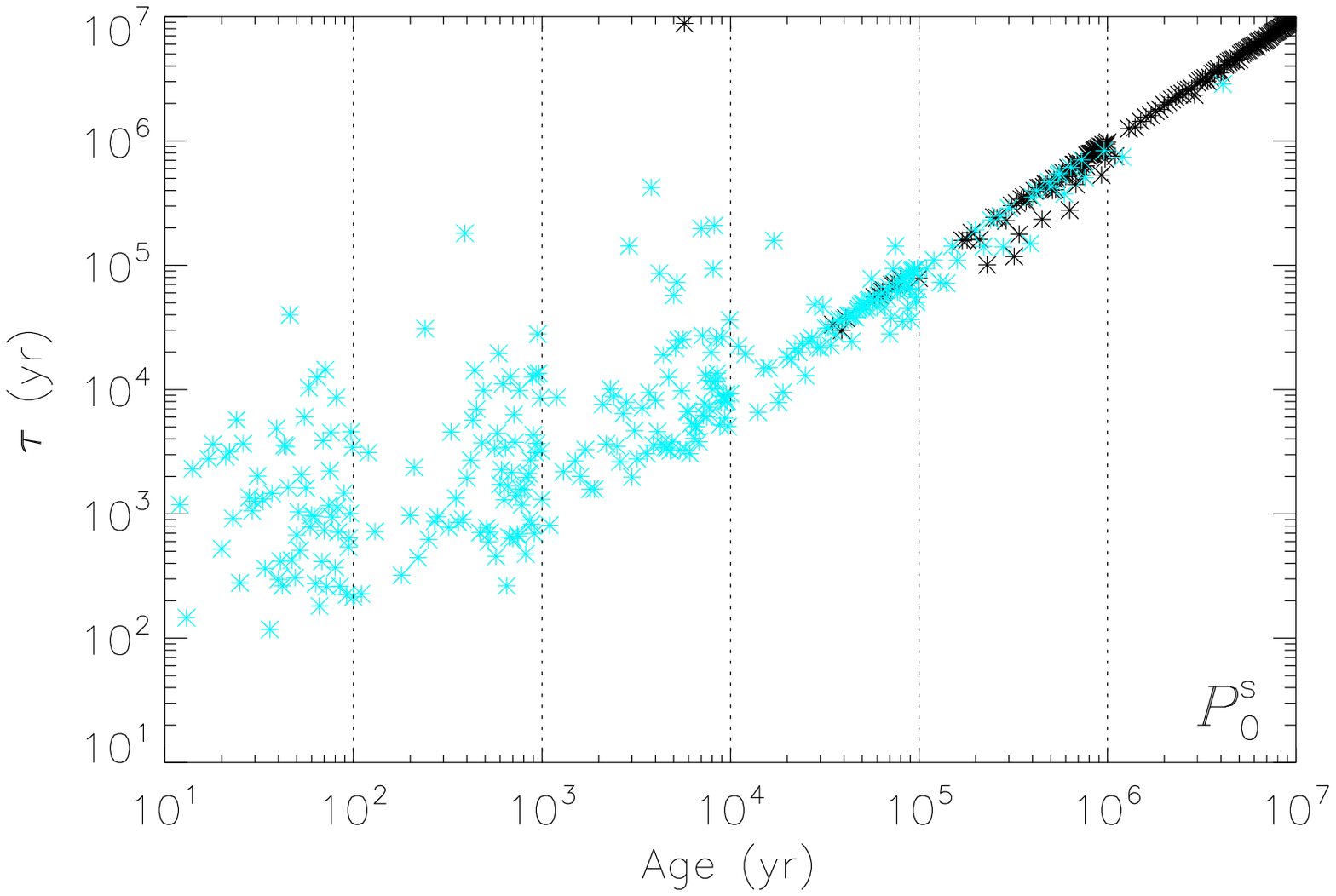}
\includegraphics[width=8.2cm]{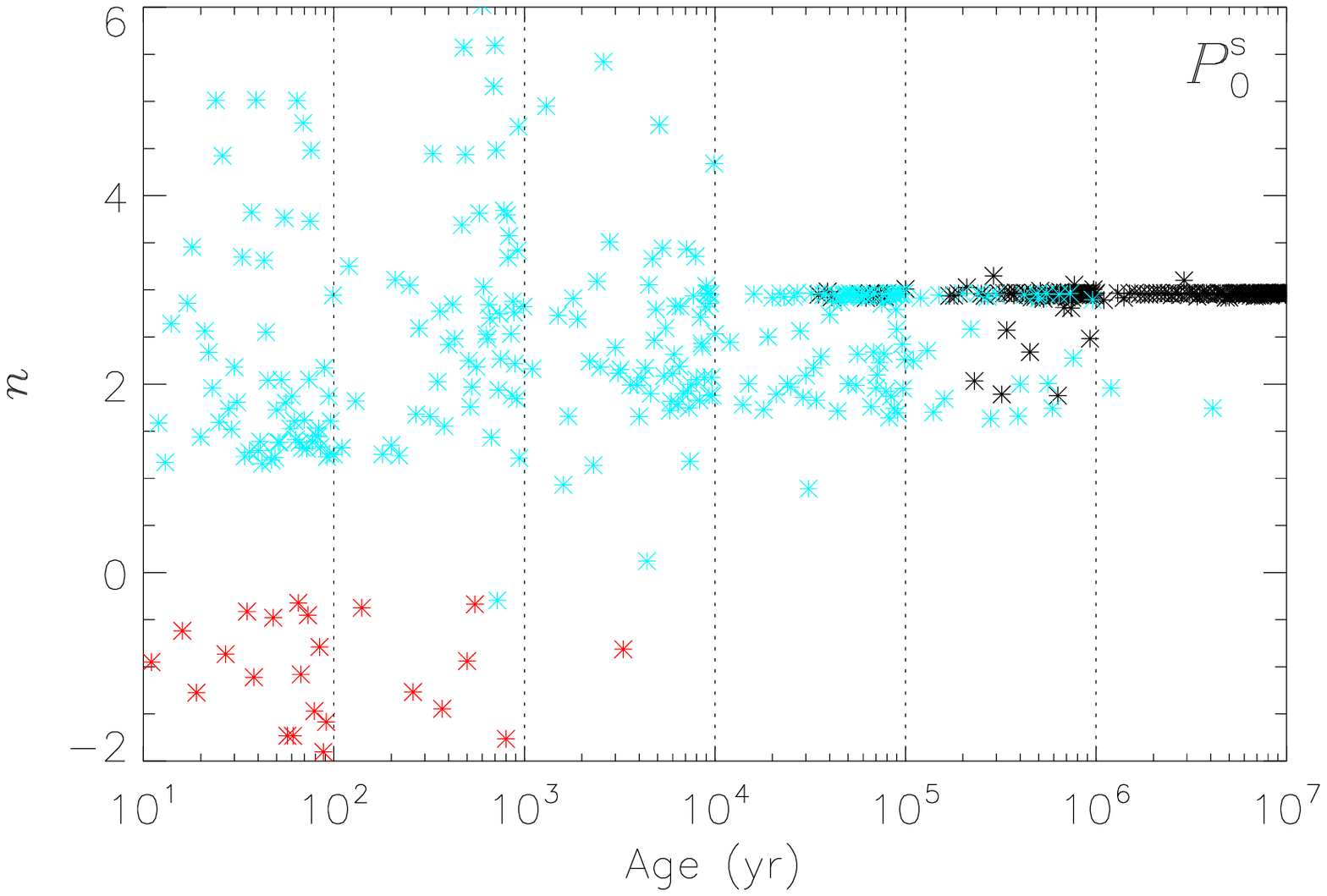}
\caption{Left-hand panels: simulated timing age distribution for 
  the fast-born (top) and slow-born (bottom) populations 
  in the presence of fallback discs. NSs 
  above the death line are plotted in cyan, and those 
  beyond the death line in black. 
  The birth rate is a function of age bin, 
  in order to have the same number of datapoints in every log bin 
  (namely, a rate of 1 yr$^{-1}$ for 10 yr $<$ Age $<$ 10$^2$ yr, 1/10
  yr$^{-1}$ for 10$^2$ yr $<$ Age $<$ 10$^3$ yr, etc.).
  Right-hand panels: same as in the left panels, but for the braking index
  distribution. NSs that are spinning up  are plotted in red.}
\label{fig:tau-and-n}
\end{figure*}

First, we examine the evolution of pulsar
periods for the initial short-period distribution 
and no death-line cutoff. We show (Fig.~\ref{fig:Pfast}, left-hand panels) 
period distributions that include all the pulsars younger than a certain age, 
for three different ages.
The top panel shows the evolution due to the combined effect of disc
and dipole radiation loss, while the bottom panel shows, for the same
initial parameters, the standard evolution due to dipole losses alone.  An
interesting and perhaps surprising feature of the evolutionary pattern
of the spin distribution in the dipole+disc case is that it becomes
bimodal. An inspection of the evolutionary tracks of individual pulsars
belonging to each peak shows that the shorter-period peak is produced 
mostly by NSs in which the dipole torque dominates over the disc torque; 
hence, the timing evolution of those pulsars is only marginally affected 
by the presence of the disc, at least when the NSs are still relatively
young. This can also be seen by comparison with the lower panel, in
which we plot the expected evolution in the absence of disc torques. 
Another (relatively small) contribution to the shorter-period peak comes 
from NSs in the disc-driven spin-up phase, which can bring their spin 
close to the initial value. The longer-period
peak of the spin distribution, on the other hand, consists of NSs
in which the disc torque dominates. We have seen (Section 2.2) that disc torques 
are almost always negative: therefore, disc-dominated NSs will more quickly 
reach longer periods. A similar period bimodality due to disc torques 
is also apparent in our simulated distribution of observable radio pulsars, 
before the death line (Fig.~\ref{fig:Pfast}, right-hand panels).

The observed period distribution of isolated pulsars 
(ATNF pulsar catalogue: Manchester at al. 2005)
is characterized by a dominant broad peak centred around a period
of several hundreds milliseconds, and another, much smaller peak,
centered around a few milliseconds. 
If fallback
discs are common around isolated NSs, then our results suggest that
the millisecond peak might correspond to the small fraction 
of the NS population which was born with the fastest periods 
and for which the disc torque was not significant. 
However, there may also be a contamination from old, weakly-magnetized 
recycled pulsars that were spun-up to millisecond periods by disc accretion 
in a binary system, but have since lost their companion star 
({\it eg}, Lorimer et al. 2004). 

Our finding that disc accretion leads to a bimodal period distribution
is particularly interesting in light of observations of NSs in
Be/X-ray binaries, which accrete material at a relatively high rate 
(from a companion, rather than from fallback).
These systems display a bimodal period distribution, with one peak
centered around $\sim 10$~s and another around $\sim 300$~s, 
previously attributed to two types of SN explosion (Knigge et al 2011). 
We propose instead that high rates of disc accretion and the competing effects 
of dipole and disc torques may have contributed to this bimodality.

We then repeated our Monte Carlo simulation for the long-period birth
distribution. We find that the period distribution 
remains single-peaked at all times (Fig.~\ref{fig:Pslow}), 
in marked contrast to the short-period birth case.
The reason is that when the initial periods are long, the
disc torque dominates over the entire life of the pulsars, for the
greatest majority of pulsars.

Fig.~\ref{fig:tau-and-n} shows the distribution of timing
ages and braking indices versus real age for the full pulsar
distribution; we have also distinguished pulsars above and below the death
line. A variable birth rate (higher for the younger objects) 
has been assumed in order to sample pulsars of different ages 
more uniformly. The fast-born pulsar distribution
shows two clusterings, one at $n\approx 3$ and another
at $n\approx 1$: the former from the younger and older pulsars, and the latter
for the intermediate phase discussed in Section 2.2. 
Note how a large fraction
of pulsars has $n<3$, but there can be some very large values. The pulsar
distribution with slow initial spins, on the other hand, is largely
concentrated in the range $1.5\lsim n\lsim 3$, plus a subset of spinning-up 
sources found predominantly with $n<0$. 
Timing ages are generally smaller than the real ages for the older objects,
and larger than the real ages for the younger NSs. 

Braking indices have been measured for a number of
pulsars. Interestingly, indices smaller than 3 are rather common, and
most often with values between 2 and 3 (e.g. Livingstone et
al. 2005a,b, 2006), with sometimes values smaller than 2 (Lyne et
al. 1996). This is the natural range of values predicted by
fallback discs, as we have shown.

\subsection{Luminosity properties}

We have seen in Section 2.1 that the initial inflow rate from 
a fallback disc can be much larger than the critical value at large radii, 
and the emitted bolometric luminosity can exceed the Eddington limit 
by a logarithmic factor. 
In our model, during this initial phase, the mass inflow at large radii 
declines as $(t/t_0)^{-19/16}$, and most of the inflow is lost in a wind 
when $\dot{M}_{\rm out} > \dot{M}_{\rm Edd}$.
We have also shown that the decay timescale is $t_0 \sim 10^2$--$10^4$ s 
for typical disc parameters. Thus, the supercritical phase 
of accretion has a very short duration.  The rapid decline in the disc 
luminosity is due to the combined effect of the decline in 
the accretion rate, and the increase of the magnetospheric radius, 
which defines the inner truncation radius of the disc.
{Note that the disc luminosity in the X-ray band, being produced in the
inner parts of the disc, is especially sensitive not only to the magnitude
of the accretion rate, but also to the spin period of the NS. The reason
lies in the fact that the inner radius of the disc, $R_{\rm in}$, being determined
by the minimum between $R_{\rm m}$ and $R_{\rm lc}$, depends on both $\dot{M}$ (through
$R_{\rm m}$), and on $\Omega$ (through $R_{\rm lc}$). The longer-wavelength emission,
on the other hand, being produced at larger disc radii, is less sensitive to
the precise location of the inner radius, and hence it has also little sensitivity
on the spin period of the NS.}

The characteristic peak disc temperature is 
\begin{eqnarray}
T_{\rm in} &\approx& \left(\frac{3GM\dot{M}}{8\pi\sigma\,R_{\rm in}^3}\right)^{1/4}\nonumber \\ 
&\approx& 
\left\{ \begin{array}{ll}
0.96 \times 10^6 \, M_{\rm NS,1}^{5/14} \, B_{12}^{-3/7} \, \dot{M}_{18}^{13/28} \ 
\mbox{K}&\\
&\hspace{-1.2cm}\mbox{for $R_{\rm in} = R_{\rm m}$}\\
1.27 \times 10^6 \, M_{\rm NS,1}^{1/4} \, \dot{M}_{18}^{1/4} \, P_{\rm 10ms}^{-3/4} \ 
\mbox{K}&\\ 
&\hspace{-1.2cm}\mbox{for $R_{\rm in} = R_{\rm lc}$}
\end{array} \right.
\end{eqnarray}
With our chosen range of initial inflow rates, we find that the peak
temperature is $\approx 10^6$ K $\approx 0.1$ keV for an initial
period of time lasting $\sim 10^2$--$10^4$ yr, and then declines $\sim
t^{-0.55}$ afterwards.  Thus, fallback discs are always much cooler
than accretion discs in NS X-ray binaries, even when they are both
near the Eddington luminosity.  This is because the magnetic field of
an old NS in an X-ray binary is {generally} weaker than in a young
pulsar\footnote{The issue of magnetic field decay (Goldreich \&
  Reisenegger 1992) is still not fully settled as far as its
  quantitative details. However, observations (e.g. see summary in
  Zhang 2007) suggest that both Atoll and Z sources (the two classes
  of low-mass X-ray binaries) have weaker $B$-fields; this
  observational evidence is based on the X-ray spectra, the luminosity
  at which they switch from propeller to accretor, and also on the
  existence of recycled millisecond pulsars, which could not be spun
  up to millisecond periods if they had a high magnetic field.}, and
the disc can extend closer to the innermost stable circular orbit or
the NS surface, reaching peak temperatures $\sim 1$--$2$ keV. This is
also the reason why disc accretion tends to spin down young pulsars
and spin up old pulsars.  In summary, we expect the direct luminosity
effect of a fallback disc to be a super-soft component in addition to
{the thermal component from the NS surface. A noteworthy result of
  our analysis is that at early times, $t\la 10^3-10^4$~yr, the
  0.1-10~keV luminosity from the disc is expected to exceed the
  luminosity from the NS surface for the range of initial accretion
  rates considered here, $\dot{M}(t=0)\ga 10^{25}$~g~s$^{-1}$.  This
  is especially so for millisecond pulsars, since at early times
  $R_{\rm in}=R_{\rm lc}$, and hence the inner radius of the disc is
  smaller for shorter-period NSs.  Lack of detection of this disc
  component in young NSs would argue against an ubiquitous formation
  of fallback discs, or that the initial accretion rates are lower
  than what numerical simulations suggest. In millisecond pulsars,
  discs could be blown away by the pulsars themselves (Eksi et
  al. 2005).}

Lastly, to conclude our discussion of the luminosity, note that,
when the non-thermal magnetospheric component is enabled,
it would likely dominate in the 2-10~keV band at least for the fastest pulsars.
In this case, 
our model suggests that the X-ray colours and spectral appearance of a
young pulsar with a fallback disc truncated at the magnetospheric
radius would be somewhat similar to those observed in the largest class of
ultraluminous X-ray sources (Feng \& Soria 2011 for a review), 
dominated by a power-law component of
photon index $\approx 2$, with a soft excess at $kT \approx
0.1$--$0.2$ keV and possibly a radiation-driven disc outflow.  This is
likely to be a coincidence due to the relative scaling of accretor
mass, accretion rate in Eddington units, and characteristic inner-disc
radius in the two classes of objects. If the thermal component of the
luminosity in both classes of objects is Eddington limited, young
pulsars with fallback discs will be one or two orders of magnitude
less luminous than ultraluminous X-ray sources with cool discs, and
there would not be possibility of misidentification if the distance is
known. Another main difference is that the X-ray luminous phase of a
fallback disc lasts only $\la 10^4$ yr, which implies that such
sources must be found very close to their natal supernova remnant;
instead, ultraluminous X-ray sources are thought to be active for $\ga
10^6$ yr and they are very rarely associated with supernova
remnants. Moreover, the characteristic inner radius of the standard
disc in power-law dominated ultraluminous X-ray sources tends to
increase with the accretion rate (for example, $R_{\rm in} \sim
\dot{M}$ if it corresponds to the spherization radius), in the
supercritical regime; as a result, the thermal component becomes
cooler but more luminous \citep{sor07,kaj09}.  Instead, the inner
radius of a disc truncated at the magnetospheric boundary will be
pushed outwards as the accretion rate declines (assuming that the NS
magnetic field varies more slowly than the accretion rate): $R_{\rm m}
\sim \dot{M}^{-2/7}$, implying $T_{\rm in} \sim \dot{M}^{13/28} \sim t^{-0.55}$ and
$L_{\rm bol} \sim T_{\rm in}^{36/13} \sim t^{-1.52}$.  At late times, the disc
becomes too faint and cool to be detectable in the X-ray band. Then the
best prospects for direct detection are at longer wavelengths, especially in the IR.

\section{Summary and Conclusions}

Fallback discs have been suggested to be a common outcome of supernova
explosions. In this paper we have explored some of the consequences of
their presence for the timing and spectral properties of the
associated NS population.  To this purpose, we have modelled the
torque and luminosity of fallback discs around young pulsars, with a
range of values for the initial spin and dipole magnetic field. {
  Relying on the results of numerical simulations of massive star
  collapse,} we assumed that the initial inflow rate of fallback
material at large radii is highly supercritical, but most of the
inflow is blown away in a radiatively driven wind, so that the
accretion rate at the inner edge of the disc is Eddington limited.

We considered two initial distributions of spin periods: one with a
Gaussian distribution centred at 5 ms, and one centred at 300 ms, with
a birth rate proportional to the SFR (assumed constant).  We evolved
the period distribution of the two populations over time, under the
combined effect of dipole and fallback disc torques. The main outcomes
of our analysis are summarized below.

\begin{enumerate}

\item We find that, within the wide range of initial parameters
  explored, the effect of a fallback disc is mostly a
  spin-down. Matter from the disc is propelled out rather than being
  accreted on the NS surface, except for a few short phases 
  when the accretion rate is
  very high.  During these phases, which can last up to
  several thousand years, the NS-disc system brightens up in
  X-rays to a luminosity that can reach the Eddington value, and the
  NS is spun up to levels that can bring the spin close to its initial
  value.

\item If fallback discs are common, they substantially
  affect our estimates of the initial spin birth of pulsars (which
  generally assume spin down losses due to dipole radiation alone). In
  particular, the period distribution of fast-born pulsars is
  predicted to evolve towards a bimodal distribution, unlike the
  distribution in the absence of fallback discs, which remains
  single-peaked if it started as such. On the other hand, for
  slow-born pulsars, the main long-term effect is simply a shift of
  the distribution to longer periods.

\item
Timing ages as measured for the NS/disc system generally
overestimate real ages for young pulsars, and underestimate it for
old pulsars. Braking indices cluster in the range $1.5\lsim
n\lsim 3$ for slow-born pulsars (apart from a short-lived spin-up phase 
with $-0.5\lsim n\lsim -2$), and $-0.5\lsim n\lsim 5$ for
fast-born pulsars. Younger objects tend to have $n\lsim 3$. 
Large values of $n$, while not common, are possible, and
associated to phase transitions in the pulsar-disc system.

\item
In addition to changing the spin evolution of a pulsar (and therefore
also its rotation-powered X-ray luminosity), a fallback disc can be
directly detected as a bright source in a broad wavelength range. We
found that at early times, the characteristic inner-disc temperature 
$kT_{\rm in} \sim 0.1$ keV for typical initial values of the magnetic field,
$B \sim 10^{12}$ Gauss, and Eddington-limited accretion, and declines
as $T_{\rm in} \sim \dot{M}^{13/28} \sim t^{-0.55}$ at later epochs.  The disc has a
low temperature compared with typical accretion discs in (older) NS
X-ray binaries, because it is truncated at the magnetospheric
radius. {The 0.1-10~keV luminosity generally exceeds that of the NS surface
at early times, if the initial accretion rates are highly superEddington.
Lack of detection of such a soft X-ray excess in young objects can help put
constraints on the ubiquity of fallback discs.}

\end{enumerate}

\section*{Acknowledgments}

RS acknowledges support from a Curtin University Senior Research Fellowship, 
and hospitality at the Mullard Space Science Laboratory (UK) and 
at the University of Sydney (Australia) during part of this work.
This work was partially supported by grants NSF
AST-1009396, AR1-12003X, and DD1-12053X (RP). {We thank an anonymous referee
for very insightful comments which greatly helped the presentation of our work.}

\end{document}